\documentclass[12pt,preprintnumbers,superscriptaddress,amsmath,amssymb,nofootinbib]{revtex4}

\usepackage{graphicx}
\usepackage{dcolumn}
\usepackage{bm}
\usepackage{amssymb}
\usepackage{amsmath}
\usepackage{epsfig}    
\usepackage{color}
\usepackage{slashed}
\usepackage{makecell}
\usepackage{float}

\begin{document} 

\title{Loop induced  $h \rightarrow \gamma \gamma, Z\gamma, gg $ decays in 3-Higgs doublet models }

%

\author{Emine Yildirim}
\email{emineyildirim@ktu.edu.tr}
\affiliation{ Department of Physics, Karadeniz Technical University,  TR61080 Trabzon, Turkey}

\begin{abstract}
\section*{Abstract}
In this paper, we analyze the loop induced decays of neutral Higgs boson $  h $
into pairs of photons, gluons, and Z$ \gamma $ in  3-Higgs Doublet Models  (3HDMs), where all the doublets have non-zero vacuum expectation values. In the 3HDM, there are contributions to these decays from loops of charged scalars $ H^{\pm}_{1,2} $ that are not present in the Standard Model.  We find an enhancement in  the channel $ h\rightarrow Z\gamma $  with respect to the Standard Model prediction in some parameter  region while respecting constraints from  the measurements of $ h\rightarrow \gamma\gamma $ at the LHC. The charged Higgs bosons $ H^{\pm}_{1,2} $ lead to the enhancement of  $ Z\gamma $  decay mode in some part of the parameter space of the 3HDM.  Then, we discuss the correlations between  the  decay modes  $ h\rightarrow Z\gamma $ and $ h\rightarrow \gamma\gamma $. Moreover, we study Type-I and Type-Z of 3HDMs and find that  these types of 3HDMs can be distinguished via only the gluonic decays, since the contributions of  the $ H^{\pm}_{1,2} $ loops  are dominant to those of  fermion loops. We also investigate the effects of the model parameters  on these loop induced decays. As an interesting phenomenological consequence,  the 3HDM is  dominantly governed by the parameters:   two ratios of vacuum expectation values (VEVs); $ \tan\gamma,~\tan\beta $ and the  mixing angles of the CP-even Higgs bosons,  $ \alpha_{1,2}~ \text{and}~\theta $.
\end{abstract}
\maketitle

\section{Introduction}

Recently a Higgs boson with a mass of 125 GeV has been discovered at the Large Hadron Collider (LHC) \cite{Aad:2012tfa, Chatrchyan:2012xdj} that resembles the Standard Model (SM) Higgs boson.
This discovery, however raises one important question, whether it is indeed the SM Higgs boson or a Higgs boson of new physics beyond the SM. 
In the class of the many beyond SM scenarios, an interesting class of Higgs-sector extension is the n-Higgs Doublet Model (nHDM).   One of the simplest but important examples is the two-Higgs-doublet model (2HDM)  with one additional scalar doublet \cite{Branco:2011iw, Bhattacharyya:2015nca}. In recent years, a lot of attention has been given to the  Three Higgs Doublet Model (3HDM) with additional two Higgs doublets \cite{my3hdm,3hdm, 3hdm1,3hdm2,3hdm3,3hdm4,3hdm5, 3hdm6,Yagyu:2016whx,Ferreira:2008zy,Machado:2010uc,  Ivanov:2012ry,GonzalezFelipe:2013xok, GonzalezFelipe:2013yhh, Keus:2013hya,  Ivanov:2014doa,Maniatis:2014oza, 3hdm-uni,Das:2019yad}.   Extended Higgs sectors naturally give rise  to Flavor-Changing Neutral Currents (FCNCs) which are in conflict with various flavour data at the tree level. This problem is usually avoided by  implementing   discrete  symmetries in  the Yukawa interactions. For example, in the 3HDM,  only one of the three doublets couples to each type of fermion.  The rich particle spectrum of the 3HDM consists of three CP-even Higgs bosons, $ h $, $ H_1 $ and $H_2$, two CP-odd Higgs bosons, $ A_1 $ and $A_2$;  and two pair of charged Higgs bosons, $H^{\pm}_1$ and $H^{\pm}_2$. In this paper, we discuss the possibility that the discovered Higgs boson at the LHC is the lightest CP-even neutral Higgs boson $h$ of the 3HDM. \\ 

In this paper, we study the loop-induced decays $h \rightarrow \gamma\gamma$,  $Z\gamma$  and  $ gg $ in the 3HDM and  their behavior over the model parameters.  The  loop induced processes are sensitive to new physics contributions, which would interfere with the SM contributions.  Although a large number of works have investigated the effects of new particles in the loop induced decay widths of the Higgs  \cite{3hdm,3hdm3,loop, loop2,loop3,Aranda:2013kq, Fontes:2014xva, Arhrib:2011vc, Hammad:2015eca}, a systematic numerical study of  the 3HDM loop induced decay widths are still missing.   In particular, we will study under which circumstances, the loop induced decay widths could be significantly affected.  We consider the possibility of constructive interference of the $ H^{\pm}_{1} $ and $ H^{\pm}_{2} $  contributions with  the SM particles contributions due to enhancement of the decay widths  $h \rightarrow  Z\gamma$   in the 3HDM. 
Besides the two ratios of Vacuum Expectation Values (VEVs), $ \tan\beta $ and $ \tan\gamma $, all loop induced decay widths are quite sensitive to parameters $ \alpha_{1,2}$ and $\theta $, which are the mixing angles between the CP even Higgs bosons. 

We will examine constraints on the parameter spaces of the  Type-I and  Type-Z- 3HDM by taking into account the measured value  of $h \rightarrow  \gamma\gamma$ decay at the LHC. Then, we will investigate whether an enhancement in the $h \rightarrow  Z\gamma$   decay mode exists while  respecting  these constraints.  We obtain an enhancement in  the $h \rightarrow  Z\gamma$   decay mode  in part  of the parameter space.   Archer-Smith \textit{et al.} recently investigated the decay rate of $h \rightarrow  Z\gamma$   and found  that simple models cannot lead to large enhancements while still respecting  other data, particularly measurements of  $h \rightarrow  \gamma\gamma$ \cite{Archer-Smith:2020gib}.   According to Ref. \cite{Chen:2013vi},  the $h \rightarrow Z \gamma$  decay rate can enhance or reduce with respect to  its SM value:   $0.76 < \Gamma (h\rightarrow Z\gamma)/ \Gamma (h\rightarrow Z\gamma)_{SM}< 2.05.$ In a study in the context of an Inert Triplet Model (ITP)  in Ref.  \cite{Kanemura:2018esc},  the $h \rightarrow Z \gamma$  decay is suppressed by about 70 \% from the SM prediction, while the $h \rightarrow \gamma \gamma$  decay remains as in the SM.  Wang \textit{et al.} investigated the $h \rightarrow \gamma \gamma$  decay in a 2HDM, but the partial decay width $\Gamma (h\rightarrow Z\gamma)$ is generally less than the SM expectation   \cite{Wang:2017xml}. 

The correlation between   $h \rightarrow  \gamma\gamma$  and $h \rightarrow  Z\gamma$ has been examined in some Refs.\cite{loop2, BhupalDev:2013xol,Chen:2013dh,Chiang:2012qz,Chen:2013vi,Chabab:2018ert}. An interesting thing  is that  correlations between the  $h \rightarrow  \gamma\gamma$  and $h \rightarrow  Z\gamma$ decays  appear in some parameter space in our model by imposing the  $h \rightarrow  \gamma\gamma$ decay  measured by the LHC.

 In particular, we  study the Type-I and Type-Z 3HDM among the five Yukawa types of 3HDMs. However, one can distinguish  these Yukawa types in in gluonic decay modes.  Only heavy quark loops contribute to the gluonic decays, while additional new charged Higgs bosons $ H^{\pm}_{1,2} $ enter the loop in the $ \gamma\gamma $ and $ Z\gamma $ decay modes as well as the SM particles. The heavy quark contributions are small compared to new charged Higgs bosons contributions  in the $ \gamma\gamma $ and $ Z\gamma $ mode, hence Yukawa types in the 3HDM could not be distinguished in these decay modes. In general, one of the most important parameters to distinguish different Yukawa types in the  3HDM is denoted as $ \tan\beta $ and $ \tan\gamma $. The gluonic and photonic decay widths are significant to investigate  the Higgs production at hadron and photon colliders, where the cross sections will be directly proportional to, respectively, the gluonic and photonic decay widths.

  Our work is organized as follows. In section \ref{2}, we briefly describe the theoretical structure of the 3HDM.  In section \ref{couplings}, we present the  couplings of the particles in the 3HDM  relevant to our calculation.  Analytic formulas needed for calculating the  decay widths  of the   $h\rightarrow \gamma\gamma,~ \gamma Z,~ gg$  are presented in  section \ref{3}, and we discuss the numerical results of the partial  decay widths and branching ratios  in the 3HDM. We finally summarize our results in section \ref{conclude}. In the appendix, we collect expressions for the loop functions used in our calculations. Moreover, details on the analysis of the scalar potential and the coefficients of the triple scalar couplings are given in the appendix.

\section{Three Higgs Doublet Model} \label{2}
\begin{table}[t]
\begin{center}
\begin{tabular}{c||c|c|c|c|c|c|c||ccc}
\hline\hline Yukawa Types &$\Phi_u$ &$\Phi_d$ &$\Phi_e$\\  \hline
Type-I    & $\Phi_2$ & $\Phi_2$ & $\Phi_2$ \\
Type-II  &  $\Phi_2$ & $\Phi_1$ & $\Phi_1$ \\
Type-X    & $\Phi_2$ & $\Phi_2$ & $\Phi_1$ \\
Type-Y   &  $\Phi_2$ & $\Phi_1$ & $\Phi_2$ \\
Type-Z    & $\Phi_2$ & $\Phi_1$ & $\Phi_3$ \\
\hline\hline
\end{tabular} 
\end{center}
\caption{The  five physically-distinct  types of Yukawa interactions  under the  $Z_2 \times \tilde{Z}_2$  symmetries  \cite{my3hdm}.
 } \label{Tab:type}
\end{table}

\begin{table}[t]
\begin{center}
\begin{tabular}{c||c|c|c||c|c|c}\hline\hline
&\multicolumn{3}{c||}{Factor for $\tilde{H}_1$, $\tilde{A}_1$ and $\tilde{H}_1^\pm$}&\multicolumn{3}{c}{Factor for  $\tilde{H}_2$, $\tilde{A}_2$ and $\tilde{H}_2^\pm$}  \\ \hline
 & $R_{u2}/R_{u1}$ & $R_{d2}/R_{d1}$ & $R_{e2}/R_{e1}$ & $R_{u3}/R_{u1}$ & $R_{d3}/R_{d1}$ & $R_{e3}/R_{e1}$          \\  \hline
Type-I       &$\cot\beta$&$\cot\beta$&$\cot\beta$&0&0&0  \\  \hline
Type-II     &$\cot\beta$&$-\tan\beta$&$-\tan\beta$&0&$-\tan\gamma/\cos\beta$&$-\tan\gamma/\cos\beta$  \\  \hline
Type-X      &$\cot\beta$&$\cot\beta$ &$-\tan\beta$&0&0&$-\tan\gamma/\cos\beta$ \\  \hline
Type-Y      &$\cot\beta$&$-\tan\beta$ &$\cot\beta$&0&$-\tan\gamma/\cos\beta$&0  \\  \hline
Type-Z      &$\cot\beta$&$-\tan\beta$  &$-\tan\beta$&0&$-\tan\gamma/\cos\beta$&$\cot\gamma/\cos\beta$  \\ 
 \hline\hline
\end{tabular} 
\end{center}
\caption{Factors appearing in Eq.~(\ref{yuk2}) for each type of Yukawa interaction. }
\label{ratios}
\end{table}

The most general Higgs potential invariant under the $SU(2)_L\times U(1)_Y\times Z_2 \times \tilde{Z}_2$ symmetry is expressed by \cite{my3hdm}
\begin{align}
V(\Phi_1,\Phi_2,\Phi_3) &= \sum_{i=1}^3 m_i^2 \Phi_i^\dagger \Phi_i - (m_{12}^2 \Phi_1^\dagger \Phi_2 
+m_{13}^2 \Phi_1^\dagger \Phi_3 
+m_{23}^2 \Phi_2^\dagger \Phi_3  + \text{h.c.}) \notag\\
 & +\frac{1}{2}\sum_{i=1}^3\lambda_i (\Phi_i^\dagger \Phi_i)^2 
+\rho_1(\Phi_1^\dagger \Phi_1)(\Phi_2^\dagger \Phi_2)
+\rho_2|\Phi_1^\dagger \Phi_2|^2 +\frac{1}{2}[\rho_3(\Phi_1^\dagger \Phi_2)^2 + \text{h.c.} ] \notag\\
&+ \sigma_1(\Phi_1^\dagger\Phi_1)(\Phi_3^\dagger\Phi_3)
 +\sigma_2|\Phi_1^\dagger \Phi_3|^2
+\frac{1}{2}[\sigma_3(\Phi_1^\dagger \Phi_3)^2 + \text{h.c.} ]\notag\\
&+ \kappa_1(\Phi_2^\dagger\Phi_2)(\Phi_3^\dagger\Phi_3)
 +\kappa_2|\Phi_2^\dagger \Phi_3|^2
+\frac{1}{2}[\kappa_3(\Phi_2^\dagger \Phi_3)^2 + \text{h.c.} ], \label{pot}
\end{align}
where  we allow the presence of the soft breaking terms  $m_{12}^2$, $m_{13}^2$ and $m_{23}^2$  for  the discrete $Z_2$ symmetries.\footnote{We use here the labelling $ m_{12},~m_{13},~m_{23} $ in place of  $ \mu_{12},~\mu_{13},~\mu_{23} $ in Ref. \cite{my3hdm}, respectively.} To avoid explicit CP violation, we take all the parameters of the scalar potential to be real.

In order to suppress  FCNCs at tree level, we will  enforce each type of fermion to couple to only a single Higgs doublet in the Yukawa sector. In the 3HDM, this condition leads to the five physically-distinct “types” given in Tab. \ref{Tab:type}. The Yukawa Lagrangian of the 3HDM reads
 
\begin{align}
-{\cal L}_Y = Y_u \bar{Q}_L (i\sigma_2)\Phi_u^* u_R^{} 
+Y_d \bar{Q}_L \Phi_d d_R^{}
+Y_e \bar{L}_L \Phi_e e_R^{} + \text{h.c.}, 
\end{align}
where $\Phi_{u,d,e}$ are either $\Phi_{1}$, $\Phi_2$ or $\Phi_3$.

The general 3HDM  contains three scalar $SU(2)_L $ doublet fields, denoted here as
\begin{align}
\Phi_i = \left[
\begin{array}{cc}
\omega_i^+ \\ 
\frac{1}{\sqrt{2}}(h_i+v_i+iz_i)
\end{array}\right],~~(i=1,...3), 
\end{align}
where the $v_i$'s are the VEVs of the $\Phi_i$'s with the sum rule $\sum_{i}v_i^2\equiv v^2 = 1/(\sqrt{2}G_F) \simeq (246$ GeV$)^2$. The VEVs may  be introduced by the two ratios
\begin{align}
\tan\beta \equiv \frac{v_2}{v_{13}}, \quad 
\tan\gamma \equiv \frac{v_3}{v_1}, ~~\text{with}~~
v_{13} \equiv \sqrt{v_1^2 + v_3^2}. 
\end{align}
Using this notation, each of the VEVs is parametrized  by 
\begin{align}
v_1 = v_{13}\cos\gamma = v\cos\beta\cos\gamma , \quad 
v_2 =  v\sin\beta,\quad
v_3 = v_{13}\sin\gamma=  v\cos\beta\sin\gamma. 
\end{align}
Scalar mass eigenstates can be obtained by rotating the fields. The rotation is performed in two stages. The first stage is to  transform all scalar fields from a weak basis with  a generic choice of vacua to the so-called Higgs basis \cite{higgsbasis,higgsbasis2}, where only 1-Higgs doublet contains the SM VEV $v$ and the Nambu-Goldstone (NG) bosons. This can be done by  rotating to the Higgs basis using the  orthogonal $3\times 3$ matrix $R$   as follows:


\begin{align}
\begin{pmatrix}
h_1 \\
h_2 \\
h_3
\end{pmatrix}=
R
\begin{pmatrix}
\tilde{h}\\
\tilde{H_1}\\
\tilde{H_2}
\end{pmatrix},~
\begin{pmatrix}
w_{1}^{\pm} \\
w^{\pm}_{2}\\
w^{\pm}_{3}
\end{pmatrix}=
R
\begin{pmatrix}
G^{\pm}\\
\tilde{H^{\pm}_{1}}\\
\tilde{H_{2}^{\pm}}
\end{pmatrix},~
\begin{pmatrix}
z_1 \\
z_2 \\
z_3
\end{pmatrix}=
R
\begin{pmatrix}
G^{0}\\
\tilde{A_1}\\
\tilde{A_2}
\end{pmatrix}. 
\end{align}

The $R$ matrix is expressed in terms of the three VEVs, following the notation of Ref. \cite{my3hdm}
\begin{align}
R &= \begin{pmatrix}
\frac{v_1}{v} &-\frac{v_2v_1}{v_{13}v} &-\frac{v_3}{v_{13}} \\
\frac{v_2}{v} &\frac{v_{13}}{v}                  & 0  \\
\frac{v_3}{v} &-\frac{v_2v_3}{v_{13}v}  &\frac{v_1}{v_{13}}
\end{pmatrix}
=
\begin{pmatrix}
\cos\gamma &0 & -\sin\gamma  \\
0&1&0 \\
\sin\gamma &0 & \cos\gamma   \\
\end{pmatrix}
\begin{pmatrix}
\cos\beta & -\sin\beta & 0\\
\sin\beta &  \cos\beta & 0\\
0&0&1 
\end{pmatrix}\notag\\
&=
\begin{pmatrix}
\cos\beta \cos\gamma & -\sin\beta\cos\gamma & -\sin\gamma\\
\sin\beta &  \cos\beta & 0\\
\cos\beta\sin\gamma&-\sin\beta\sin\gamma&\cos\gamma 
\end{pmatrix}.
\end{align}\label{R}

After spontaneous symmetry breaking (SSB), the  NG bosons $G^\pm$ and $G^0$  are absorbed into the longitudinal components of the  $W^{\pm}$ and $Z$ respectively. In Eq.~(6), two singly-charged states $\tilde{H}_a^\pm$, two CP-odd states $\tilde{A}_a$ and three CP-even states $\tilde{h}$ and $\tilde{H}_a$
are not given in the  mass eigenstates. The next stage is  to determine the  scalar mass eigenstates through the orthogonal transformations
\begin{align}
\begin{pmatrix}
G^{\pm}\\
H^{\pm}_{1}\\
H_{2}^{\pm}
\end{pmatrix}=
O_{c}^{\dagger}
\begin{pmatrix}
w^{\pm}_{1}\\
w^{\pm}_{2}\\
w^{\pm}_{3}
\end{pmatrix},~
O_{\theta_{2}}=
\begin{pmatrix}
 1 & 0 & 0 \\
 0 & \cos (\theta_2 ) & -\sin (\theta_2 ) \\
 0 & \sin (\theta_2 ) & \cos (\theta_2 ) \\
\end{pmatrix}
\end{align}
\begin{align}
\begin{pmatrix}
G^{0}\\
A_{1}\\
A_{2}
\end{pmatrix}=
O^{\dagger}_{A}
\begin{pmatrix}
z_{1}\\
z_{2}\\
z_{3}
\end{pmatrix},~
O_{\theta_{1}}=
\begin{pmatrix}
 1 & 0 & 0 \\
 0 & \cos (\theta_1 ) & -\sin (\theta_1 ) \\
 0 & \sin (\theta_1) & \cos (\theta_1 ) \\
\end{pmatrix},
\end{align}
where  $ \theta_{2} $ and $ \theta_{1} $ are the mixing angles for the charged  and CP-odd states, respectively.  $O_c= O_{\theta_2}R,~~O_A= O_{\theta_1}R.$ 

For the CP-even states, we can obtain the mass eigenstates  through
\begin{align}
\begin{pmatrix}
\tilde{h}_1  \\
\tilde{H}_1 \\
 \tilde{H}_2  
\end{pmatrix} = O^h
\begin{pmatrix}
h  \\
H_1 \\
H_2  
\end{pmatrix}.
\end{align}
where $O^h= R^T O$ and
\begin{align}
O=R_{23}(\theta)R_{12}(\alpha_1) R_{13}(\alpha_2)\label{O}
\end{align}
where  $\alpha_{1,2},~\theta$  mixing angles  in the CP-even sector.
\begin{align}
R_{13}(\alpha_2)=
\begin{pmatrix}
\cos\alpha_2 & -\sin\alpha_2 &0  \\
\sin\alpha_2 & \cos\alpha_2 &0 \\
0&0&1  
\end{pmatrix},~
R_{12}(\alpha_1)=
\begin{pmatrix}
\cos\alpha_1 & 0 & -\sin\alpha_1  \\
0&1&0 \\
\sin\alpha_1 &0& \cos\alpha_1  
\end{pmatrix},~
R_{23}(\theta)=
\begin{pmatrix}
 1 & 0 & 0 \\
 0 & \cos (\theta ) & -\sin (\theta ) \\
 0 & \sin (\theta ) & \cos (\theta ) \\
\end{pmatrix}.
\end{align}
The masses of the three neutral physical states, $h,$ $ H_1$ and $H_2$,  are evaluated in detail  in   Appendix \ref{appendix-cp even}. Consequently, the model has nine scalar physical particles: (i) three CP-even scalars $(h, H_{1,2}),$ (ii) two CP-odd scalars $(A_{1,2}),$ and (iii) four charged scalars$ (H^\pm_{1,2}).$ In this paper, we focus on  the physics related to the  Higgs bosons $h, ~ H_{1},~ H_{2} $, $H_1^\pm$ and $H_2^\pm$, for which there are 
eighteen significant parameters,  namely
\begin{align}
m_h,~ m_{H_1},~m_{H_2},~m_{H_1^\pm}^{},~m_{H_2^\pm}^{},~m_{A_{1}},~m_{A_{2}} , ~v,~\tan\beta,~\tan\gamma,~\theta_1,~ \theta_2,  ~\alpha_{1,2},~\theta, m_{12},~m_{23},~\text{and}~m_{13}. \notag 
\end{align}

\subsection{The Yukawa sector}

The Yukawa couplings of the  Higgs bosons are given by \cite{my3hdm}

\begin{align}
&-{\cal L}_Y = \sum_{f=u,d,e}\frac{m_{f^i}}{v}\bar{f}^i\left[\left(\tilde{h} + \frac{R_{f2}}{R_{f1}}\tilde{H}_1 +  \frac{R_{f3}}{R_{f1}}\tilde{H_2} \right) 
-2I_f\left(\frac{R_{f2}}{R_{f1}}\tilde{A}_1 +  \frac{R_{f3}}{R_{f1}}\tilde{A_2} \right) \gamma_5  \right]f^i \notag\\
&+\frac{\sqrt{2}}{v}\left[\bar{u}^j V_{ji} m_{d^i} \left(\frac{R_{d2}}{R_{d1}}\tilde{H}_1^+ +  \frac{R_{d3}}{R_{d1}}\tilde{H}_2^+ \right)P_R  d^i  
-\bar{u}^im_{u^i} V_{ij} \left(\frac{R_{u2}}{R_{u1}}\tilde{H}_1^+ +  \frac{R_{u3}}{R_{u1}}\tilde{H}_2^+ \right)P_L  d^j\right]  + \text{h.c.}\notag\\
&+\frac{\sqrt{2}}{v}\bar{\nu}^i  m_{e^i} \left(\frac{R_{e2}}{R_{e1}}\tilde{H}_1^+ + \frac{R_{e3}}{R_{e1}}\tilde{H}_2^+ \right)P_R  e^i   + \text{h.c.} \label{yuk2}
\end{align}
where $I_f=+1/2\,(-1/2)$ for $f=u\,(d,e)$ and $V_{ij}$ is the Cabibbo-Kobayashi-Maskawa (CKM) matrix element. 
The ratios of the matrix elements $R_{f2}/R_{f1}$ and 
$R_{f3}/R_{f1}$ ($f=u,d,e$) are collected in Tab.~\ref{ratios} for   each of the five types of Yukawa interactions. 
\subsection{Kinetic Lagrangian}
It is useful to take a closer look at kinetic Lagrangian to extract the trilinear scalar-gauge-gauge couplings

\begin{align}\label{kineticlag}
&{\cal L}_{kin} = \sum_{i=1}^{3} | D_{\mu}\Phi_{i}|^{2}\ni   \dfrac{g^{2}}{2} W^{+}_{\mu}W^{\mu-}\left(   \sum_{i=1}^{3}  \nu_{i} h_{i} \right) \\ \nonumber
&=\dfrac{g^{2}\nu}{2} W^{+}_{\mu}W^{-\mu}\left(  \frac{1}{\nu} \sum_{i=1}^{3}  \nu_{i} h_{i} \right)         
\end{align}
where g represents the $SU(2) _{L} $ gauge coupling.
\section{Relavant COUPLINGS AND ANALYTIC FORMULAS} \label{couplings}
\subsection{Scalar couplings}
 The scalar, $ hH^{+}_{i} H^{-}_{i}~(i=1,2)$,  trilinear couplings, which are relevant to the decay widths discussed in the  next section, could be expressed through  $ \beta, \gamma,\alpha_{1,2}, \theta$ and other parameters as follows:
\begin{align}
&\lambda_{hH_{1}^{+}H_{1}^{-}}=a ~m_{h_{1}}^2+b~ m_{H_{1}^{\pm}}^2 +c~ m_{H_{2}^{\pm}}^2 +d ~m_{12}^{2}+e ~m_{13}^{2}+f~ m_{23}^{2} \\
&\lambda_{hH_{2}^{+}H_{2}^{-}}=a_{1} ~m_{h_{1}}^2+b_{1}~ m_{H_{1}^{\pm}}^2 +c_{1}~ m_{H_{2}^{\pm}}^2  +d_1 ~m_{12}^{2}+e_1 ~m_{13}^{2}+f_1~ m_{23}^{2} 
\end{align}
 $a, b, c, d, e, f,  a_1, b_1, c_1,d_1, e_1, f_1$, which are coefficients of $ h H^{+}_{i} H^{-}_{i}$ terms in the potential, are expressed explicitly in Appendix  \ref{app-coeff}.
 
\subsection{Yukawa couplings }
To calculate the Yukawa couplings of the neutral Higgs bosons, we define the  coupling coefficients  $B_{1,2,3}$   in terms of the elements of the  matrix R given  in Tab. \ref{ratios} \cite{my3hdm}.
In this case, the Yukawa couplings of the neutral Higgs boson takes the following form:
\begin{align}
&-{\cal L}_Y =  \sum_{f=u,d,e}\frac{m_f}{v}\bar{f}\left[\left (h  B_{1}+ H_1 B_{2}+ H_2 B_{3}\right]\right ) f,     \label{yuk2a}
\end{align}
we find 
\begin{align}
B_1 &=c_{\alpha _1} c_{\alpha _2} c_{\beta } c_{\gamma }-\frac{c_{\alpha _1} c_{\alpha _2} c_{\gamma } R_{\text{f2}} s_{\beta }}{R_{\text{f1}}}-\frac{c_{\alpha _1} c_{\alpha _2} R_{\text{f3}} s_{\gamma }}{R_{\text{f1}}}+\frac{c_{\alpha _2} c_{\gamma } c_{\theta } R_{\text{f3}} s_{\alpha _1}}{R_{\text{f1}}}+c_{\alpha _2} c_{\beta } c_{\theta } s_{\alpha _1} s_{\gamma }-\\  \notag
&\frac{c_{\alpha _2} c_{\theta } R_{\text{f2}} s_{\alpha _1} s_{\beta } s_{\gamma }}{R_{\text{f1}}}-\frac{c_{\alpha _2} c_{\beta } R_{\text{f2}} s_{\alpha _1} s_{\theta }}{R_{\text{f1}}}-c_{\alpha _2} s_{\alpha _1} s_{\beta } s_{\theta }+\frac{c_{\beta } c_{\theta } R_{\text{f2}} s_{\alpha _2}}{R_{\text{f1}}}+\\ \notag
&\frac{c_{\gamma } R_{\text{f3}} s_{\alpha _2} s_{\theta }}{R_{\text{f1}}}+c_{\beta } s_{\alpha _2} s_{\gamma } s_{\theta }+c_{\theta } s_{\alpha _2} s_{\beta }-\frac{R_{\text{f2}} s_{\alpha _2} s_{\beta } s_{\gamma } s_{\theta }}{R_{\text{f1}}}\\
B_2&=\frac{c_{\alpha _2} c_{\beta } c_{\theta } R_{\text{f2}}}{R_{\text{f1}}}+c_{\alpha _2} c_{\theta } s_{\beta }+\frac{c_{\alpha _2} c_{\gamma } R_{\text{f3}} s_{\theta }}{R_{\text{f1}}}+c_{\alpha _2} c_{\beta } s_{\gamma } s_{\theta }-\frac{c_{\alpha _2} R_{\text{f2}} s_{\beta } s_{\gamma } s_{\theta }}{R_{\text{f1}}}-c_{\alpha _1} c_{\beta } c_{\gamma } s_{\alpha _2}+\\ \notag
&\frac{c_{\alpha _1} c_{\gamma } R_{\text{f2}} s_{\alpha _2} s_{\beta }}{R_{\text{f1}}}+\frac{c_{\alpha _1} R_{\text{f3}} s_{\alpha _2} s_{\gamma }}{R_{\text{f1}}}-\frac{c_{\gamma } c_{\theta } R_{\text{f3}} s_{\alpha _1} s_{\alpha _2}}{R_{\text{f1}}}-\\ \notag
&\frac{c_{\theta } R_{\text{f2}} s_{\alpha _1} s_{\alpha _2} s_{\beta } s_{\gamma }}{R_{\text{f1}}}+\frac{c_{\beta } R_{\text{f2}} s_{\alpha _1} s_{\alpha _2} s_{\theta }}{R_{\text{f1}}}+c_{\beta } c_{\theta } s_{\alpha _1} s_{\alpha _2} s_{\gamma }+s_{\alpha _1} s_{\alpha _2} s_{\beta } s_{\theta }\\
B_3&=\frac{c_{\alpha _1} c_{\gamma } c_{\theta } R_{\text{f3}}}{R_{\text{f1}}}+c_{\alpha _1} c_{\beta } c_{\theta } s_{\gamma }-\frac{c_{\alpha _1} c_{\theta } R_{\text{f2}} s_{\beta } s_{\gamma }}{R_{\text{f1}}}-\frac{c_{\alpha _1} c_{\beta } R_{\text{f2}} s_{\theta }}{R_{\text{f1}}}+\\ \notag
&\frac{c_{\gamma } R_{\text{f2}} s_{\alpha _1} s_{\beta }}{R_{\text{f1}}}-c_{\beta } c_{\gamma } s_{\alpha _1}-c_{\alpha _1} s_{\beta } s_{\theta }+\frac{R_{\text{f3}} s_{\alpha _1} s_{\gamma }}{R_{\text{f1}}},
\end{align}
where $c_x^{}$, $s_x^{}$ and $t_x $ are shorthand notations for 
$\cos x$, $\sin x$ and $ \tan x$, respectively.\\
These $ B_{1,2,3} $ coefficients  stand for the couplings $ g_{hff},~g_{H_1ff} $ and $ g_{H_2ff} $, receptively.

\subsection{Scalar-gauge-gauge couplings}
 The gauge$-$gauge$-$scalar type interactions are extracted from Eq.(14) as
\begin{align}
h W^+{}_{\mu } W^{\mu -} &=\frac{1}{2} g^2 v \left(c_{\beta } \left(c_{\alpha _1} c_{\alpha _2} c_{\gamma }+s_{\gamma } \left(c_{\alpha _2} c_{\theta } s_{\alpha _1}+s_{\alpha _2} s_{\theta }\right)\right)+s_{\beta } \left(c_{\theta } s_{\alpha _2}-c_{\alpha _2} s_{\alpha _1} s_{\theta }\right)\right)
\end{align}

\section{LOOP INDUCED DECAYS INTO $   \gamma\gamma,~\gamma Z,~gg $}\label{3}
In the 3HDM,  the  decay widths of the  $ h \rightarrow \gamma\gamma, Z\gamma$ receive  contributions from the new charged scalars $ H^{\pm}_{1,2} $.  The  decay widths of  $ h$ into $ \gamma\gamma,~\gamma Z~ \text{and} ~ gg$   are given explicitly by \cite{djouadi,djouadi1, hunter} 
\begin{align}
\Gamma (h\rightarrow \gamma\gamma)= &\frac{G_F \alpha^2 m_h^3}{128 \sqrt{2}\pi^3} \mid\sum_f N_c Q_f^2 g_{hff}A_{1/2}(\tau_f) + g_{h WW} A_1(\tau_W)\\ \nonumber 
&+\frac{m_W^2 \lambda_{hH^{+}_{i} H^{-}_{i}}}{2c_w^2 m_{H^{\pm}_i}}A_0(\tau_{H^{\pm}_i}) \mid^2 ,\\ 
\Gamma (h \rightarrow Z\gamma)= &\frac{G_F^2  m_W^2\alpha m_h^3}{64\pi^4} \left( 1-\frac{m_Z^2}{m_h^2}\right) \mid\sum_f   g_{h ff} \frac{N_c Q_f \hat v_f}{c_w}A_{1/2}(\tau_f, \lambda_f)\\ \nonumber
& +g_{h WW} A_1(\tau_W, \lambda_W) +\frac{m_W^2 v_{H^{\pm}_i} }{2c_w^2 m_{H^{\pm}_i}}\lambda_{h H^{+}_{i} H^{-}_{i}}A_0(\tau_{H^{\pm}_i})  \mid^2 \\
\Gamma (h\rightarrow gg)&= \frac{G_F \alpha_{s}^{2} m^{3}_{h}}{36\sqrt{2}\pi^3}\mid \frac{3}{4}\sum_f g_{hf f}A_{1/2}(\tau_f)\mid^2
\end{align}
with $ \hat v_f= 2 I_f^3 -4Q_f s^2_W $ and $ v_{H ^{\pm}_i}=\frac{2c_W^2-1}{c_W}$.

Here $ N_c $ is the color factor $N_c=3(1)$ for quarks(leptons) and  $Q_f$ stands for the electric charge  of a particle  in the loop.  $ \alpha $ and $ \alpha_s $ are  the fine-structure constant and strong coupling constant, respectively. Dimensionless one-loop  factors $A_1$ (for the W boson), $ A_{1/2} $ (for the fermions, $f=t,b,\tau$) and $ A_0 $ (for the charged scalars, $ H^{\pm}_{1,2} $ ) are written in Appendix \ref{loop function}. The couplings $ g_{h ff} $ and $ g_{h W^{+}W^{-}} $ and $ \lambda_{h H^{+}_{i} H^{-}_{i}} $ are expressed in Section \ref{couplings}. \\

The decay $ h \rightarrow Z\gamma $ in the 3HDM receives  contributions from the new charged scalars $ H^{\pm}_{1,2} $, hence its decay width could be  enhanced in some parameter regions.

\subsection{  $ h \rightarrow \gamma\gamma$}  \label{gam-sec}


We now consider the decay width of $h \rightarrow \gamma \gamma$ in the Type-I  3HDM.  We first discuss   constraints from the LHC  measurements of the decay  of $h \rightarrow \gamma \gamma$ (Ref. \cite{ ATLAS:2018hxb})  in our model parameter spaces. By using the measured value of  $h \rightarrow \gamma \gamma$  \cite{ATLAS:2018hxb}, we perform an extensive scan within the parameter space of the model. We then report the allowed values  of the model parameters in Table \ref{limit} to apply  in all figures in subsection B.

\begin{table}[!t]
\resizebox{\columnwidth}{!}{
\scalebox{0.12}{
\begin{tabular}{ll}
\begin{tabular}{|c|c|c|c|c|c|}\hline
$ \tan\gamma $ &  $ \tan\beta=1 $ &$ \tan\beta=2 $& $ \tan\beta=3 $\\ \hline

$ \tan\gamma=1 $ & \makecell{$ \alpha_{1,2}=\theta=\frac{\pi}{3}$  \\
$ m_{H_{1}^{\pm}} =205,300~ \text{GeV}$}&\makecell{$ \alpha_{1,2}=\theta=\frac{\pi}{3}$  \\ 
$ m_{H_{1}^{\pm}} =319, 500 ~\text{GeV}$}&\makecell{$ \alpha_{1,2}=\theta=\frac{\pi}{3}$  \\
$ m_{H_{1}^{\pm}} =457,760 ~\text{GeV}$} \\ \hline

$ \tan\gamma=1 $ & \makecell{$ \alpha_{1,2}=\theta=\frac{13\pi}{45}, \frac{14\pi}{45},\frac{\pi}{3}, \frac{67\pi}{90} $\\
$ m_{H_{1}^{\pm}} =300 ~\text{GeV}$}&\makecell{$ \alpha_{1,2}=\theta=\frac{\pi}{3}, \frac{4\pi}{9},\frac{41\pi}{90}$  \\ 
$ m_{H_{1}^{\pm}} =300 ~\text{GeV}$}&\makecell{} \\ \hline

$ \tan\gamma=2 $ & \makecell{$ \alpha_{1,2}=\theta=\frac{\pi}{9}, \frac{4\pi}{9}$  \\
$ m_{H_{1}^{\pm}} =290 ~\text{GeV}$}&\makecell{}&\makecell{} \\ \hline
$ \tan\gamma=3 $ & \makecell{$ \alpha_{1,2}=\theta=\frac{23\pi}{90}, \frac{42\pi}{90}, \frac{49\pi}{90}$  \\
$ m_{H_{1}^{\pm}} =320 ~\text{GeV}$}&\makecell{}&\makecell{} \\ \hline

\end{tabular}
\end{tabular}
}}
\vspace{5mm}
\caption{Allowed values of the parameters  in a Type-I 3HDM obtained by  performing the measured value of $ h\rightarrow \gamma\gamma $ decay at the LHC. Here we consider $ m_{h}=125$ GeV, $\theta_{2}= \pi/6$, $ m_{H_{2}^{\pm}}= m_{H_{1}^{\pm}}+100$ GeV and $ m_{12}=m_{23}=m_{13}=200 $ GeV. 
}
\label{limit}
\end{table}

\begin{figure}[t]
\begin{center}
\includegraphics[width=52mm]{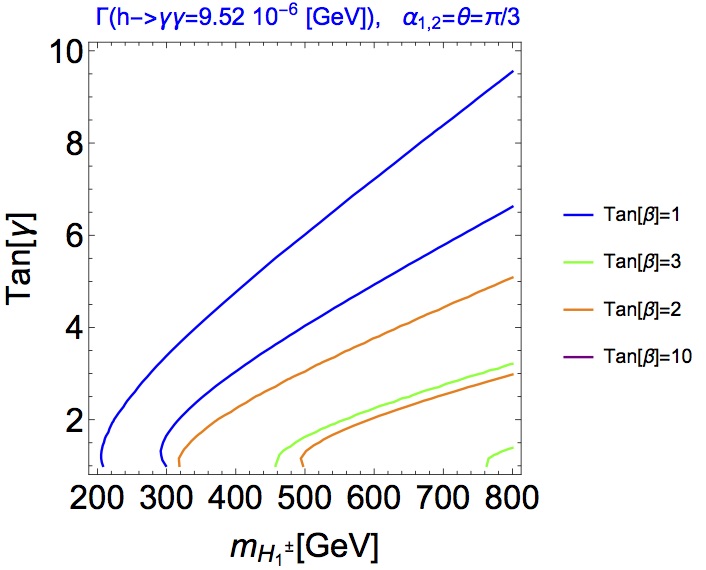}\hspace{1.0mm}
\includegraphics[width=52mm]{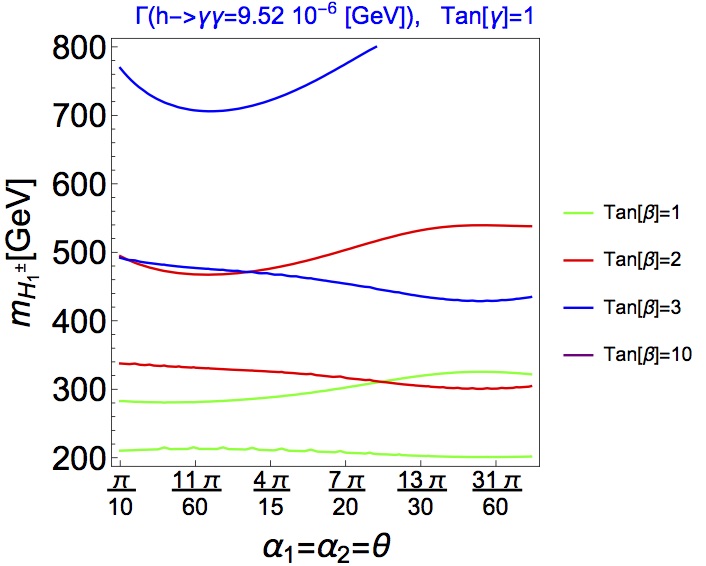}\hspace{1.0mm}
\includegraphics[width=52mm]{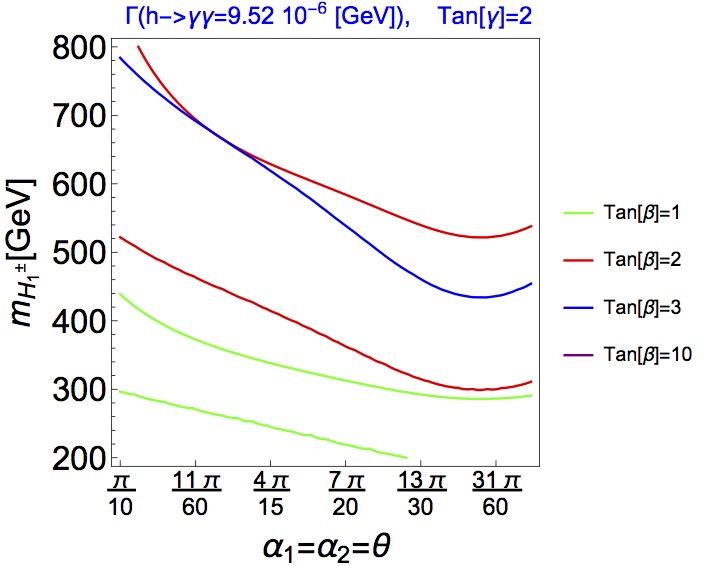}\hspace{1.0mm}
\caption{
 The contour lines show the allowed  points of parameter spaces from the  measurements of $ h \rightarrow \gamma \gamma$ decay  at the LHC in the  left plane $(\tan\gamma,~ m_{H_{1}^{\pm}} )$   and  center and right planes $ ( m_{H_{1}^{\pm}},~ \alpha_{1,2}=\theta )$  in the Type-I   3HDM.  
The other parameters are fixed to be:  $ m_{h}=125$ GeV, $\theta_{2}= \pi/6$, $ m_{H_{2}^{\pm}}= m_{H_{1}^{\pm}}+100$ GeV and $ m_{12}=m_{23}=m_{13}=200 $ GeV. 
}
\label{plotlhc}
\end{center}
\end{figure}

In Fig.~\ref{plotlhc},  contour  plots in the ($ \tan\gamma, m_{H^{\pm}_{1}}) $ and $(m_{H^{\pm}_{1}}, \alpha_{1,2} =\theta$ ) planes show the partial decay width for $ h\rightarrow \gamma\gamma $. In the plots, the allowed  regions are bounded  by the contour lines by applying  the  measured values of Higgs to diphotons at the LHC. We observed from the left panel of  Fig.~\ref{plotlhc} that the experimental constraint rules out the region of  $ \tan\gamma \geq 10$ for the cases of $ \alpha_{1,2}=\theta=\frac{\pi}{3}$ and $ \tan\beta \geq 1$.  The left panel shows that there are allowed regions of the parameter space in the range of  $200~ \text{GeV} \leq m _{H^{\pm}_{1,2}} \leq 750 ~\text{GeV} $ for the cases of $ \tan\gamma=1$ and $ 1\leq \tan\beta \leq 3 $. It is clear from the center and right panels of Fig.~\ref{plotlhc} that there are allowed parameter spaces for the cases of $\frac{\pi}{10} \leq\alpha_{1,2},~\theta \leq \frac{13\pi}{30}, $   $ \tan\gamma=1,2 $ and $ 1\leq \tan\beta\leq  3$. We can see that allowed regions depend heavily upon  the choice of $ \tan\beta,~\tan\gamma,~\text{and}~m_{H_{1,2}^{\pm}} $.  In addition, our scan shows that the curves in  Fig. ~\ref{plotlhc}  are  not  much sensitive with the changes of  $\alpha_{1,2}, \theta$ for the case of $ \tan\gamma=1 $. We have fixed the mixing angle  $ \theta_{2}=\pi/6 $ in all figures.

\subsection{  $ h \rightarrow Z\gamma$}

\begin{figure}[t]
\begin{center}
\includegraphics[width=76mm]{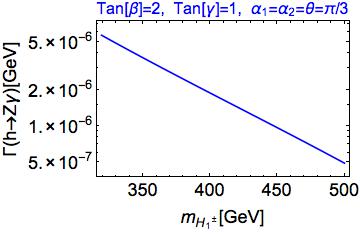}\hspace{4.0mm}
\includegraphics[width=76mm]{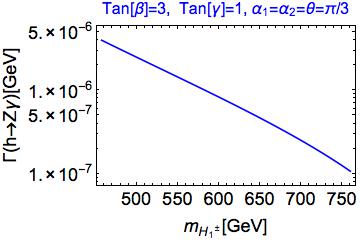}\hspace{4.0mm}
\vspace{2.0mm}
\includegraphics[width=76mm]{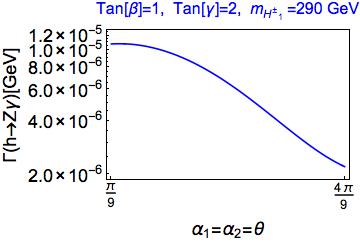}\hspace{4.0mm}
\includegraphics[width=76mm]{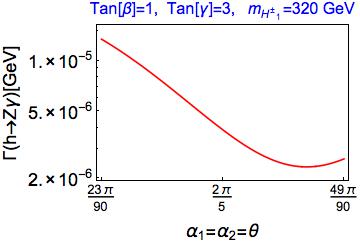}\hspace{4.0mm}
\caption{The decay width of $h \rightarrow Z \gamma$ 
as a function of $m_{H^{\pm}_{1}}$(upper panel) and  $\alpha_{1,2}=\theta$ (lower panel) in the Type-I   3HDM.  
We take $ m_{H_{2}^{\pm}}= m_{H_{1}^{\pm}}+100$ GeV, $ m_{h}=125 $ GeV,  $ m_{12}=m_{23}=m_{13}=200 $ GeV, and  $ \theta_{2}=\pi/6 .$
}
\label{mixh}
\end{center}
\end{figure}

In this section, we will examine the decay partial width of $ h\rightarrow Z\gamma$ while respecting the parameter spaces that are not ruled out by the  $ \Gamma(h\rightarrow \gamma\gamma) $ analysis. Using the  values of parameters in Table \ref{limit}, we first focus on the decay width of $ \Gamma( h\rightarrow Z\gamma)  $ in the Type I 3HDM. The dependence of $ \Gamma (h\rightarrow Z\gamma) $  on $ m_{H_{1}^{\pm}} $ is shown on the upper panels of Fig. \ref{mixh} for $ \tan\gamma=1 $ and different choice of $ \tan\beta$. The curves show that the decay width is  sensitive to the  charged Higgs bosons masses.   One can see that the decay width drops sharply with the charged Higgs masses $ m_{H^{\pm}_{1,2}} $ increasing.  The decay width  is sensitive to the choice of $ \tan\beta $. On the lower panel of   Fig. \ref{mixh}, we interpret the dependence of the decay with on the CP-even mixing angles $\alpha_{1,2},~\theta$  for various values of $ \tan\gamma$ and  $ m_{H_{1}^{\pm}}.  $ We see that in order to have large values of $h \rightarrow Z\gamma$   small values of  the CP-even mixing angles are required.

\begin{figure}[t]
\begin{center}
\includegraphics[width=70mm]{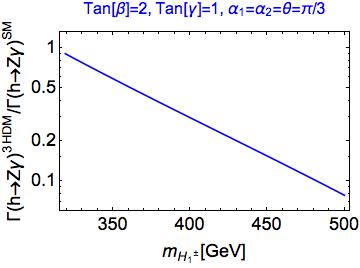}\hspace{4.0mm}
\includegraphics[width=70mm]{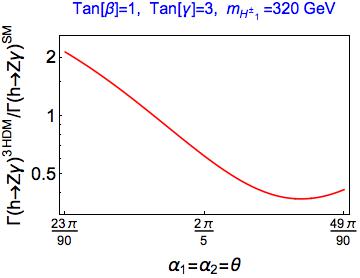}\hspace{4.0mm}
\caption{Dependence of  
 $ \Gamma(h\rightarrow Z \gamma)_{\text{3HDM}}/ \Gamma(h \rightarrow Z \gamma)_{\text{SM}}$ on 
   $m_{H_{1}^{\pm}}$(left panel)  and   $\alpha_{1,2},~\theta$ (right panel) in the Type-I   3HDM.  
We take $m_{H_{1}^{\pm}}= m_{H_{2}^{\pm}}= 100$ GeV ,  $ m_{h}=125 $ GeV,  $\theta_{2}= \pi/6$ and $ m_{12}=m_{23}=m_{13}=200 $ GeV. }
\label{mixhz}
\end{center}
\end{figure}

In  Fig. \ref{mixhz}, we present the deviation in  the partial decay width of  $h \rightarrow Z\gamma$ relative to the SM prediction.   On the left panel  range $ 319\leq m_{H_{1}^{\pm}}\leq 500 $ is plotted, and on the right panel the range  $ 23\pi/90\leq \alpha_{1,2}=\theta \leq 49\pi/90 $  is plotted.  From these plots, it is possible to see that the deviation in  $\Gamma(h \rightarrow Z\gamma)$ increases with large values of  $ m_{H^{\pm}_{1,2}} $ and  $ \alpha_{1,2}=\theta $. On the right panel, the partial decay width of $h \rightarrow Z\gamma$ with respect to the SM one enhances about by a factor 2 for the case of $ \tan\gamma=3 $ and $   \alpha_{1,2}=\theta=23\pi/90.  $  As can be seen from the left panel, for 
$ m_{H_{1}^{\pm}}=319 $ GeV, the deviation from the SM prediction is about 10 \%  when $ \tan\beta=2 $ and $ \tan\gamma=1 $.

To compare the branching ratio of BR$( h\rightarrow Z \gamma )$ in  the Type I 3HDM to the  branching ratio   of BR$( h\rightarrow Z \gamma )$ in the SM,  we define the ratio of  two branching ratios
\begin{align}
R_{\text{BR}}= \frac{ BR(h\rightarrow Z\gamma)^{\text{3HDM}}}{BR(h\rightarrow Z\gamma)^{\text{SM}}},
\end{align}\label{BR-eq}
where BR$ (h \rightarrow Z\gamma)^{\text{SM} }$ represents the branching ratio in the SM.
\begin{figure}[t]
\begin{center}
\includegraphics[width=70mm]{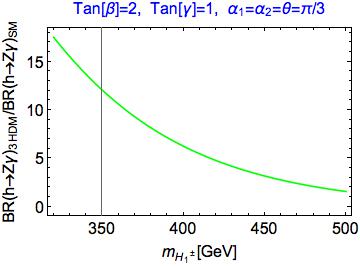}\hspace{4.0mm}
\includegraphics[width=70mm]{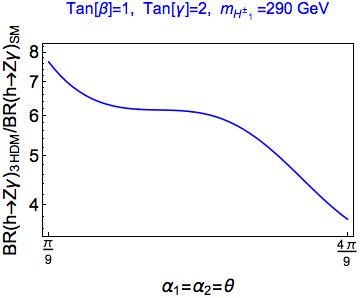}\hspace{4.0mm}
\caption{Dependence of  
 $\text{BR}(h\rightarrow Z \gamma)_{\text{3HDM}}/ \text{BR}(h \rightarrow Z \gamma)_{\text{SM}}$ on 
   $m_{H_{1}^{\pm}}$ (left panel) and  $\alpha_{1,2},~\theta$ (right panel)  in the Type-I   3HDM.  
We take $m_{H_{1}^{\pm}}= m_{H_{2}^{\pm}}+100$ GeV, $m_{h}=125 $ GeV,  $\theta_{2}= \pi/6$ and $ m_{12}=m_{23}=m_{13}=200 $ GeV. }
\label{BR}
\end{center}
\end{figure}
 Fig. \ref{BR} shows the deviation in the branching ratio of  BR$(h\rightarrow Z\gamma)^{\text{3HDM}}  $ with respect to  the SM prediction  as a function of $m_{H_{1}^{\pm}}$  in the Type I-3HDM. $ R_{BR}>1 $ occurs due to the large contribution of the $ H^{\pm}_{1,2} $ loop.  The left plot  in Fig. \ref{BR}  presents a deviation from the SM prediction at  small values of $ m_{H^{\pm}_{1} }$  and $ \tan\beta= 2.$  The right plot  in Fig. \ref{BR}   appears to be compatible  with the SM prediction for $ \tan\beta=1,  $  $ \tan\gamma= 3$ and  the large values of $ \alpha_{1,2}=\theta$.    As can be seen from the left plot, BR$(h\rightarrow Z\gamma)^{\text{3HDM}}  $  is close to the SM prediction at values of $ m_{H_{1}^{\pm}}=500 $ GeV. The current searches at the LHC have a limit of about 4 times the $h\rightarrow Z \gamma  $  signal strength in the Standard Model \cite{ATLAS:2020qcv}. Thus, this decay would be observed in the high-luminosity (HL) run of the LHC  when   $  R_{BR} \simeq 1 $.  It is clear from the left and right plots that the enhancement of the  $ h\rightarrow Z\gamma $  branching ratio is so large  in some regions of parameter space hence the HL-LHC  could have the potential to exclude this parameter region.

Note that the ratio of two branching ratios BR$(h\rightarrow Z\gamma)^{\text{3HDM}}/\text{BR}(h\rightarrow Z\gamma)^{\text{SM}}$ is larger than that of two partial widths $ \Gamma(h\rightarrow Z \gamma)^{\text{3HDM}}/ \Gamma(h \rightarrow Z \gamma)^{\text{SM}}$. 
In the Type-I 3HDM, the total decay  width  is smaller than   the SM prediction   because the  partial widths of $( h\rightarrow \bar{f}f,~gg )$ are significantly suppressed with respect to their  SM values.  As a result, there is an enhancement  in the  BR$(h\rightarrow Z\gamma)^{\text{3HDM}}$ relative to its SM value. 

We note that we have explicitly checked that the partial  decay widths of $(h\rightarrow Z\gamma ( \gamma\gamma))$ in the  Type-I 3HDM are  nearly the same as  those in the Type-Z 3HDM.

\begin{align}
R_{Z\gamma (\gamma\gamma)}= \frac{ \Gamma(h\rightarrow Z\gamma   (\gamma\gamma))^{\text{3HDM}}}{\Gamma(h\rightarrow Z\gamma (\gamma\gamma))^{\text{SM}}},
\end{align}\label{gam-eq}
where $\Gamma (h \rightarrow Z\gamma)^{\text{SM} }$ represents the partial width in the SM.

\begin{figure}[t]
\begin{center}
\includegraphics[width=86mm]{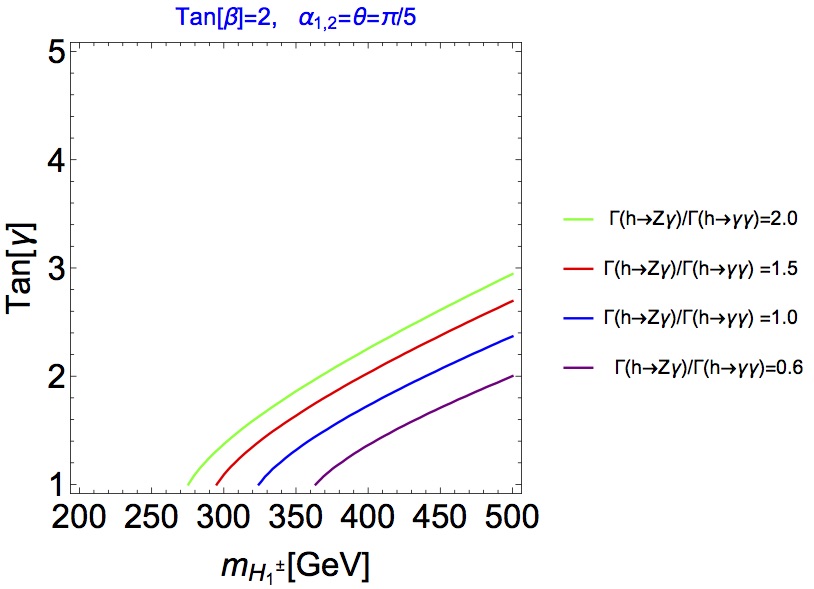}\hspace{4.0mm}
\includegraphics[width=70mm]{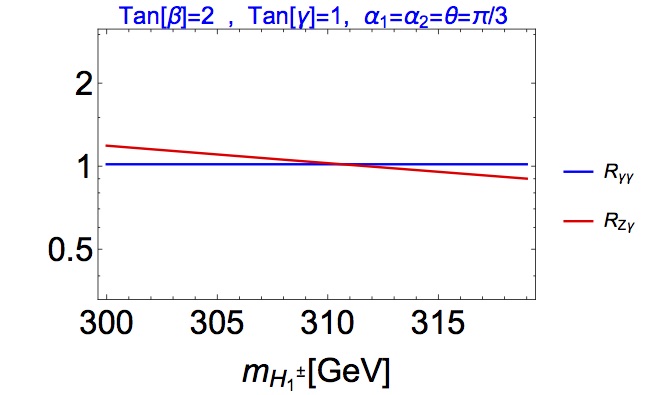}\hspace{4.0mm}
\caption{Left plane shows contour plot of  $ \Gamma^{\gamma\gamma}/ \Gamma^{Z\gamma}$ as a function of $m_{H^{\pm}_{1}}$ and $\tan\gamma $   in the Type-I   3HDM.   Right plane shows  correlation between $ R_{\gamma\gamma} $ and  $ R_{Z\gamma} $    in the Type-I   3HDM.  
We take $ m_{h}=125 $ GeV, $m_{H_{1}^{\pm}}= m_{H_{2}^{\pm}}+100$ GeV, $\theta_{2}= \pi/6$ and   $ m_{12}=m_{23}=m_{13}=200 $ GeV .}
\label{cor3}
\end{center}
\end{figure}

Before concluding, remarks on  the correlations between the $ h\rightarrow \gamma\gamma $ and  $ h\rightarrow Z \gamma $  partial widths  in  Type-I 3HDM are in order. The left panel of Fig. \ref{cor3} shows the contour plot of  $ \Gamma^{\gamma\gamma}/ \Gamma^{Z\gamma}$ in the plane of $ (m_{H^{\pm}_{1}},\tan\gamma)$    with the constraint from the measurement $h\rightarrow \gamma\gamma$ decay. One can see that the allowed regions increase with large value of $ \Gamma^{\gamma\gamma}/ \Gamma^{Z\gamma}$. 
 The left panel shows   the   regions where  the partial width of $h\rightarrow Z\gamma$  could be  either larger or smaller than that of $h\rightarrow \gamma\gamma $. On the right panel of Fig. \ref{cor3}, we discuss the correlations between $ R_{\gamma\gamma} $ and  $ R_{Z\gamma}, $ defined in Eq. (26),  while we retain the decay of  $ h\rightarrow \gamma\gamma  $ as in its measured value at the LHC.   The right panel illustrates that the correlation is sensitive to the choice of  $ m_{H^{\pm}_{1}} $.   The plots in Fig. \ref{cor3} show that  correlation  can exist between   the  $h\rightarrow Z\gamma$ and $h\rightarrow \gamma\gamma$  decays at particular parameter regions. This correlation could be tested in  future experiments at the LHC.

\subsection{  $ h \rightarrow gg$}
\begin{figure}[t]
\begin{center}
\includegraphics[width=76mm]{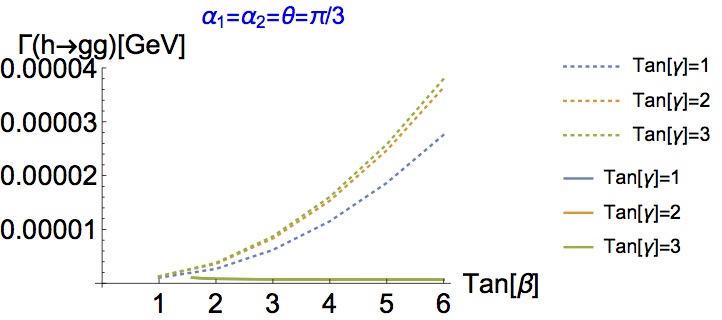}\hspace{4.0mm}
\includegraphics[width=76mm]{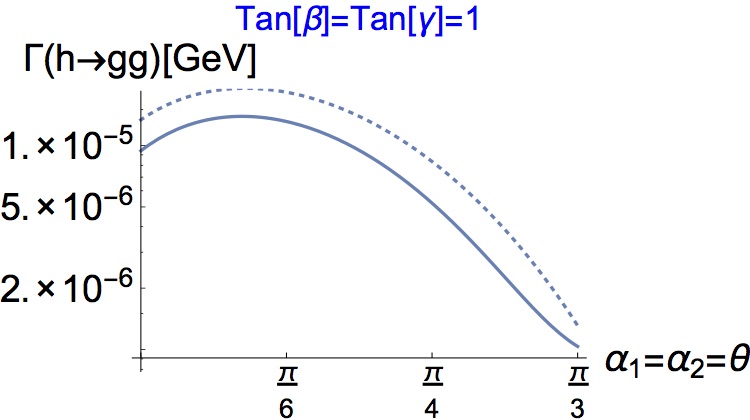}\hspace{4.0mm}
\caption{The left plot indicates the  decay width of $h\rightarrow gg$ 
as a function of $\tan\beta$ in the Type-I (solid lines) and  Type-Z  (dotted lines) 3HDM.  The right plot indicates the  decay width of $h\rightarrow gg$ 
as a function of the CP-even mixing angles in the Type-I (solid lines) and  Type-Z  (dotted lines) 3HDM. 
We take   $m_h=125 $ GeV.
}
\label{hgg}
\end{center}
\end{figure}

In this section,  we discuss the gluonic  decay widths of the  125 GeV  Higgs boson  $h$.
First, we present the decay width $ \Gamma(h\rightarrow gg) $ as a function of $ \tan\beta $ for various values of $ \tan\gamma $ on the left plot of Fig.\ref{hgg}.  The solid and dotted lines stand for the Type-I  and  Type-Z 3HDM respectively. Hence one can use them to distinguish  the Type-I and  Type-Z 3HDM. The decay of $h$ into $gg$ increases with $    \tan\gamma $ and $ \tan\beta $ in the Type-Z 3HDM, while it remains the same at large $\tan\beta  $ in the Type-I 3HDM. 

The right plot of Fig. \ref{hgg} indicates the dependence of  the CP-even mixing angles on the decay of  $ h $ into $ gg $.  The decay width is very sensitive to the change of $\alpha_{1,2}~ \text{and}~\theta$.  The solid and dotted lines stand for the Type-I  and Type-Z 3HDM, respectively, and thus  one can disentangle the two types in the gluonic decay.\\

In general, some of the  important parameters to distinguish different Yukawa types of the model are denoted as $ \tan\beta~\text{and} ~\tan\gamma.$

\section{CONCLUSIONS} \label{conclude}
In this work, we have analyzed the new physics contributions predicted by the  3HDM  to the loop induced decays $ h\rightarrow \gamma \gamma, ~\gamma Z,~ gg $. We have found that 

\begin{itemize}
 \item {The new charged Higgs bosons $H^{\pm}_{1,2}$ can significantly alter the  decay widths $  h \rightarrow \gamma \gamma, ~\gamma Z$  (where h is the lightest CP-even scalar in the 3HDM),  since their loop factor is the largest among others.}
  \item {tan$ \beta $, tan$ \gamma $, $ \alpha_{1,2},~\theta $, and  $ m_{H_{1,2}^{\pm}}$ parameters can have significant effects on these decays.}
  
\item{We can disentangle  the Type-I  and Type-Z 3HDM only in the gluonic decays, since   $ \gamma\gamma $ and $Z \gamma  $ decays  are dominated by the $ H^{\pm}_{1} $ and  $ H^{\pm}_{2} $ bosons loop contributions.}
 
  \item{ $ h\rightarrow Z \gamma $ branching ratio increases considerably compared to its SM value in the part of the parameter space while the $h\rightarrow \gamma\gamma $ decay is compatible with the LHC measurement.  Our results on the BR$( h\rightarrow Z \gamma )$   can be tested at the LHC when $R_{BR} \simeq 1 $. }

 
 
 \item{A correlation  between  the $ h\rightarrow Z \gamma $  and  $ h\rightarrow \gamma \gamma $ decays  exist, which could be probed in  future experiments at the LHC.} 
 
\end{itemize}

In general, a large enhancement of $ \Gamma(h\rightarrow Z\gamma) $ in the 3HDM may  be obtained with the contributions of the new charged Higgs bosons $H^{\pm}_{1,2}$ and probed at the HL-LHC.

\begin{appendix}

\newpage 
\section{Mass Matrices} 
Now, we determine the mass matrices in different sectors.
\subsection{CP charged scalar sector}
 The analytic expressions for the mass matrices of the singly-charged 
(${\cal M}_\pm^2$) scalar states in the Higgs basis of 3HDMs  can be extracted from the scalar potential as
\begin{align}
V_C=\begin{pmatrix}
w_1^+  & w_2^+  & w_3^+
\end{pmatrix}
M_\pm^2
\begin{pmatrix}
w_1^-  \\
 w_2^-  \\ w_3^-
\end{pmatrix}.
\end{align}
The (${\cal M}_\pm^2$)  $3 \times 3$ mass matrix can be diagonalized as follows:
\begin{align}
(M_C)^2=R^T\cdot M_\pm^2 \cdot R=
\begin{pmatrix}
0 &  0 &0 \\
0 & (M_C)^2_{22}  & (M_C)^2_{23} \\
0 & (M_C)^2_{23}  & (M_C)^2_{33}   
\end{pmatrix}.
\end{align}

The matrix elements of ${\cal M}_C^2$ are given by 

\begin{align}
({\cal M}_C^2)_{22} &= -\frac{v^2}{2}\left[(\rho_2 + \rho_3)c_\gamma^2 + (\kappa_2 + \kappa_3)s_\gamma^2)   \right] 
+ m_{12}^2\frac{c_\gamma}{s_\beta c_\beta}+m_{23}^2\frac{s_\gamma}{s_\beta c_\beta}  , \\
({\cal M}_C^2)_{33} &= -\frac{v^2}{2}\left[(\rho_2 + \rho_3)s_\beta^2s_\gamma^2 + (\sigma_2 + \sigma_3)c_\beta^2 + (\kappa_2 + \kappa_3)s_\beta^2c_\gamma^2   \right]\notag\\
&\quad +m_{12}^2t_\beta s_\gamma t_\gamma + \frac{m_{13}^2}{s_\gamma c_\gamma} +m_{23}^2 \frac{t_\beta c_\gamma}{ t_\gamma} , \\
({\cal M}_C^2)_{23} &= -\frac{v^2}{4}(\rho_2 + \rho_3 -\kappa_2 -\kappa_3)s_\beta s_{2\gamma}  
+ m_{12}^2\frac{s_\gamma}{c_\beta}-m_{23}^2\frac{c_\gamma}{c_\beta}. 
\end{align}\label{charged1}
This matrix can be fully diagonalized by employing the orthogonal matrix $O_{\theta_{2}}$ as follows:
\begin{align}
O_{\theta_{2}}.(M_C)^2 . O_{\theta_{2}}^T=
\begin{pmatrix}
0 &  0 &0 \\
0 & m^{2}_{H^\pm_1}  & 0 \\
0 & 0  &  m^{2}_{H^\pm_2}
\end{pmatrix}
~~~~~~
O_{\theta_{2}}=
\begin{pmatrix}
 1 & 0 & 0 \\
 0 & \cos (\theta_2 ) & -\sin (\theta_2 ) \\
 0 & \sin (\theta_2 ) & \cos (\theta_2 ) \\
\end{pmatrix}
\end{align} \label{charged}

Using Eqs. (A3)$-$ (A6), $\kappa_2, \sigma_2~ \text{and}~ \rho_2$ can be solved as
\begin{align}
\kappa_2&=\frac{2 m_{A_1}^2 c_{\theta _1}^2-\frac{2 m_{A_2}^2 c_{\theta _1} s_{\theta _1}}{s_{\beta } t_{\gamma }}+2 m_{A_2}^2 s_{\theta _1}^2+\frac{m_{A_1}^2 s_{2 \theta _1}}{s_{\beta } t_{\gamma }}-4 c_{\theta _2}^2 m_{H_1{}^\pm}^2+\frac{m_{H_2{}^\pm}^2 s_{\theta _2}}{c_{\gamma } s_{\beta } s_{\gamma }}-4 m_{H_2{}^\pm}^2 s_{\theta _2}^2+\frac{m_{H_2{}^\pm}^2 s_{\theta _2}}{s_{\beta } t_{\gamma }}}{2 v^2}+\\ \notag
&\frac{-\frac{2 c_{\theta _2} s_{\theta _2} t_{\gamma } \left(\frac{m_{H_1{}^\pm}^2}{s_{\gamma }^2}+\frac{m_{H_1{}^\pm}^2}{t_{\gamma }^2}+m_{H_2{}^\pm}^2\right)}{s_{\beta }}+\frac{m _{12}^2 c_{\gamma }}{c_{\beta } s_{\beta }}+\frac{m _{23}^2 s_{\gamma }}{c_{\beta } s_{\beta }}+\frac{m _{23}^2}{c_{\beta } s_{\beta } s_{\gamma }}-\frac{m _{12}^2}{c_{\beta } c_{\gamma } s_{\beta }}+\frac{m _{12}^2 s_{\gamma } t_{\gamma }}{c_{\beta } s_{\beta }}+\frac{m _{23}^2 c_{\gamma }}{c_{\beta } s_{\beta } t_{\gamma }}+\frac{(m_{H_1{^\pm}}^2) s_{\theta _2} t_{\gamma }}{s_{\beta }}}{2 v^2}\\
\rho_2&= \frac{1}{2 v^2}\left(2 m_{A_1}^2 c_{\theta _1}^2-4 c_{\theta _2}^2 m_{H_1{}^\pm}^2+2 m_{A_2}^2 s_{\theta _1}^2+\frac{m_{H_1{}^\pm}^2 s_{\theta _2}}{c_{\gamma } s_{\beta } s_{\gamma }}-4 m_{H_2{}^\pm}^2 s_{\theta _2}^2+\frac{m_{H_2{}^\pm}^2 s_{\theta _2}}{s_{\beta } t_{\gamma }}-\frac{m_{A_1}^2 s_{2 \theta _1} t_{\gamma }}{s_{\beta }}+\right. \notag \\
&\frac{m_{A_2}^2 s_{2 \theta _1} t_{\gamma }}{s_{\beta }}+\frac{m_{H_1{}^\pm}^2 s_{\theta _2} t_{\gamma }}{s_{\beta }}-\frac{2 c_{\theta _2} s_{\theta _2} t_{\gamma } \left(m_{H_2{}^\pm}^2 \left(\frac{1}{s_{\gamma }^2}+1\right)+\frac{m_{H_1{}^\pm}^2}{t_{\gamma }^2}\right)}{s_{\beta }}+ \\ \notag
&\left.\frac{m _{12}^2 c_{\gamma }}{c_{\beta } s_{\beta }}+\frac{m _{12}^2}{c_{\beta } c_{\gamma } s_{\beta }}+\frac{m _{23}^2 s_{\gamma }}{c_{\beta } s_{\beta }}-\frac{m _{23}^2}{c_{\beta } s_{\beta } s_{\gamma }}+\frac{m _{12}^2 s_{\gamma } t_{\gamma }}{c_{\beta } s_{\beta }}+\frac{m _{23}^2 c_{\gamma }}{c_{\beta } s_{\beta } t_{\gamma }}+\frac{m_{H_1{}^\pm}^2 s_{\theta _2}^2 t_{\gamma }^2 \left(\frac{m_{H_1{}^\pm}^2}{s_{\gamma }^2}+\frac{m_{H_1{}^\pm}^2}{t_{\gamma }^2}+m_{H_2{}^\pm}^2\right)}{s_{\beta }^2}\right)\\ 
\sigma_2&= \frac{1}{4 v^2}\left(\frac{\left(m_{A_1}^2-m_{A_2}^2\right) \left(c_{2 \beta }-3\right) c_{\theta _1}^2}{c_{\beta }^2}-\frac{2 \left(m_{A_1}^2-m_{A_2}^2\right) c_{2 \gamma } s_{2 \theta _1} t_{\beta }}{c_{\beta } c_{\gamma } s_{\gamma }}+\frac{4 c_{2 \gamma } \left(m_{H_1{}^\pm}^2-m_{H_2{^\pm}}^2\right) s_{\theta _2} t_{\beta }}{c_{\beta } c_{\gamma } s_{\gamma }}+\right.\notag \\
&\left(t_{\beta }^2-1\right) \left(\left(m_{A_1}^2-m_{A_2}^2\right) s_{\theta _1}^2-m_{A_1}^2-m_{A_2}^2+c_{2 \theta } \left(m_{H_1{}^\pm}^2-m_{H_2{}^\pm}^2\right)+m_{H_1{}^\pm}^2+3 m_{H_2{}^\pm}^2\right)+\\ \notag
&\frac{3 m_{A_1}^2 s_{\theta _1}^2+m_{A_1}^2+m_{A_2}^2-2 \left(c_{2 \beta }-3\right) c_{\theta _2}^2 \left(m_{H_1{}^\pm}^2-m_{H_2{}^\pm}^2\right)-2 m_{H_1{}^\pm}^2-2 m_{H_2{}^\pm}^2}{c_{\beta }^2}\\ \notag
&\frac{3 m_{A_2}^2 s_{\theta _1}^2-\frac{4 m _{13}^2}{c_{\gamma } s_{\gamma }}-6s_{\theta _2}^2 m_{H_1{}^\pm}^2 +6s_{\theta _2}^2 m_{H_2{}^\pm}^2 +m _{13}^2 s_{2 \gamma }+2 m _{13}^2 s_{\gamma }^2 t_{\gamma }+6 m _{13}^2 t_{\gamma }+\frac{8 m _{13}^2}{t_{\gamma }}}{c_{\beta }^2}\Biggr).
\end{align}

\subsection{CP odd scalar sector}  
The analytic expressions for the mass matrices of the CP-odd scalars
$({\cal M}_P^2)$  in the Higgs basis of 3HDM  can be extracted from the scalar potential as
\begin{align}
V_C=\begin{pmatrix}
z_1  &z_2  & z_3
\end{pmatrix}
\frac{M_P^2}{2}
\begin{pmatrix}
z_1  \\
z_2 \\ z_3
\end{pmatrix}.
\end{align}
The CP-odd scalar  mass matrix  ${\cal M}_P^2$ can be diagonalized as follows:
\begin{align*}
(M_A)^2=R^T\cdot M_P^2 \cdot R=
\begin{pmatrix}
0 &  0 &0 \\
0 & (M_A)^2_{22}  & (M_A)^2_{23} \\
0 & (M_A)^2_{23}  & (M_A)^2_{33}   
\end{pmatrix}.
\end{align*}
The matrix elements of   ${\cal M}_A^2$ is given by 

\begin{align}
({\cal M}_A^2)_{22} &= -v^2\left(\rho_3 c_\gamma^2 +\kappa_3 s_\gamma^2\right) +  m_{13}^2\frac{c_\gamma}{s_\beta c_\beta} + m_{23}^2\frac{s_\gamma}{s_\beta c_\beta}, \\ 
({\cal M}_A^2)_{33} &= -v^2\left(\rho_3 s_\beta^2s_\gamma^2 +\sigma_3 c_\beta^2 +\kappa_3 s_\beta^2 c_\gamma^2 \right) 
+m_{12}^2 t_\beta s_\gamma t_\gamma + m_{13}^2 \frac{s_\gamma}{s_\gamma c_\gamma} + m_{23}^2 \frac{t_\beta c_\gamma}{t_\gamma} , \\ 
({\cal M}_A^2)_{23} &= -\frac{v^2}{2}(\rho_3-\kappa_3)s_\beta s_{2\gamma} + m_{12}^2\frac{s_\gamma}{c_\beta} - m_{23}^2\frac{c_\gamma}{c_\beta}.
\end{align}
In order to   fully  diagonalise this matrix,  an orthogonal matrix  $O_{\theta_{1}}$ can be employed as
\begin{align}
O_{\theta_{1}}.(M_A)^2 . O_{\theta_{1}}^T=
\begin{pmatrix}
0 &  0 &0 \\
0 & m^{2}_{A_1}  & 0 \\
0 & 0  &  m^{2}_{A_2}
\end{pmatrix}
~~~~~~
O_{\theta_{1}}=
\begin{pmatrix}
 1 & 0 & 0 \\
 0 & \cos (\theta_1 ) & -\sin (\theta_1 ) \\
 0 & \sin (\theta_1 ) & \cos (\theta_1 ) \\
\end{pmatrix}.
\end{align} \label{charged}
Using Eqs. (A11)$- $(A14), $\kappa_3, \sigma_3~ \text{and}~ \rho_3$ can be solved as
\begin{align}
\kappa_3=&  \frac{-2 m_{A_1}^2 c_{\theta _1}^2+\frac{2 \left(m_{A_2}^2-m_{A_1}^2\right) c_{\theta _1} s_{\theta _1}}{s_{\beta } t_{\gamma }}-2 m_{A_2}^2 s_{\theta _1}^2+\frac{m _{23}^2 s_{\gamma }}{c_{\beta } s_{\beta }}+\frac{m _{23}^2}{c_{\beta } s_{\beta } s_{\gamma }}-\frac{m _{12}^2}{c_{\beta } c_{\gamma } s_{\beta }}+\frac{m _{12}^2 s_{\gamma } t_{\gamma }}{c_{\beta } s_{\beta }}+\frac{m _{23}^2 c_{\gamma }+m _{12}^2 s_{\gamma }}{c_{\beta } s_{\beta } t_{\gamma }}}{2 v^2}\\
\rho_3=&\frac{-2 m_{A_1}^2 c_{\theta _1}^2-2 m_{A_2}^2 s_{\theta _1}^2+\frac{\left(m_{A_1}^2-m_{A_2}^2\right) s_{2 \theta _1} t_{\gamma }}{s_{\beta }}+\frac{m _{12}^2}{c_{\beta } c_{\gamma } s_{\beta }}+\frac{m _{23}^2 s_{\gamma }}{c_{\beta } s_{\beta }}-\frac{m _{23}^2}{c_{\beta } s_{\beta } s_{\gamma }}+\frac{m _{12}^2 s_{\gamma } t_{\gamma }}{c_{\beta } s_{\beta }}+\frac{m _{23}^2 c_{\gamma }+m _{12}^2 s_{\gamma }}{c_{\beta } s_{\beta } t_{\gamma }}}{2 v^2} \\
\sigma_3=&-\frac{\frac{c_{2 \beta } \left(\left(m_{A_1}^2-m_{A_2}^2\right) c_{2 \theta _1}+m_{A_1}^2+m_{A_2}^2\right)-3 \left(m_{A_1}^2-m_{A_2}^2\right) c_{\theta _1}^2+3 m_{A_1}^2 s_{\theta _1}^2-3 m_{A_2}^2 s_{\theta _1}^2+m_{A_1}^2+m_{A_2}^2-\frac{4 m _{13}^2}{c_{\gamma } s_{\gamma }}}{c_{\beta }}+\frac{2 \left(m_{A_2}^2-m_{A_1}^2\right) c_{2 \gamma } s_{2 \theta _1} t_{\beta }}{c_{\gamma } s_{\gamma }}}{4 v^2 c_{\beta }} 
\end{align}
\subsection{CP even scalar sector} \label{appendix-cp even}
In the general weak basis, the CP-even scalar mass matrix $M^2_H$ can be extracted from the potential as

\begin{align}
V_S=\begin{pmatrix}
h_1 & h_2  & h_3
\end{pmatrix}
\frac{M_H^2}{2}
\begin{pmatrix}
h_1  \\
 h_2  \\ 
 h_3
\end{pmatrix}.
\end{align}
The matrix elements of  ${\cal M}_H^2$ are given by 
\begin{align}
({\cal M}_H^2)_{11} &=\frac{m_{12}^2 t_{\beta }}{c_{\gamma }}+\lambda _1 v^2 c_{\beta }^2 c_{\gamma }^2+m_{13}^2 t_{\gamma }, \\
({\cal M}_H^2)_{22} &= \frac{m_{12}^2 c_{\gamma }}{t_{\beta }}+\frac{m_{23}^2 s_{\gamma }}{t_{\beta }}+\lambda _2 v^2 s_{\beta }^2,\\
({\cal M}_H^2)_{33} &=\lambda _3 v^2 c_{\beta }^2 s_{\gamma }^2+\frac{m_{23}^2 t_{\beta }}{s_{\gamma }}+\frac{m_{13}^2}{t_{\gamma }}, \\
({\cal M}_H^2)_{12} & =\left(\rho _1+\rho _2+\rho _3\right) v^2 c_{\beta } c_{\gamma } s_{\beta }-m_{12}^2, \\
({\cal M}_H^2)_{13} & =\left(\sigma _1+\sigma _2+\sigma _3\right) v^2 c_{\beta }^2 c_{\gamma } s_{\gamma }-m_{13}^2, \\
({\cal M}_H^2)_{23} & = \left(\kappa _1+\kappa _2+\kappa _3\right) v^2 c_{\beta } s_{\beta } s_{\gamma }-m_{23}^2. 
\end{align}
In the above expressions, we used the shorthand notations $c_X^{}=\cos X$, $s_X^{}=\sin X$ and $t_X = \tan X$.  

For the masses of the three neutral physical states, $h,$ $ H_1$ and $H_2$,  we need to diagonalise the $3\times 3$ mass matrix $M^2_H$, defined in the general weak basis, by employing the orthogonal matrix $O$ given in Eq.~(\ref{O}) as 

\begin{align}
O^T\cdot M_H^2 \cdot O=
\begin{pmatrix}
m_h^2 &  0 &0 \\
0 & m_{H_1}^2  & 0 \\
0 & 0  & m_{H_2}^2   
\end{pmatrix}.
\end{align}
Finally,  the remaining parameters  $\lambda_1, ~\lambda_2, ~\lambda_3, ~\kappa_1, \sigma_1 ~ \text{and} ~ \rho_1 $  can be obtained from Eqs.  (A19)$-$(A25) as follows:

\begin{align}
\lambda_1&=-\frac{c_{\alpha _1}^2 \left(-\left(c_{\alpha _2}^2 m_h^2+m_{H_1}^2 s_{\alpha _2}^2\right)\right)+\frac{m _{12}^2 t_{\beta }}{c_{\gamma }}-m_{H_2}^2 s_{\alpha _1}^2+m _{13}^2 t_{\gamma }}{v^2 c_{\beta }^2 c_{\gamma }^2}  \\
\lambda_2&=\frac{c_{\alpha _2}^2 \left(c_{\theta }^2 m_{H_1}^2+m_h^2 s_{\alpha _1}^2 s_{\theta }^2\right)-c_{\alpha _2} s_{\alpha _1} s_{\alpha _2} s_{2 \theta } \left(m_h^2-m_{H_1}^2\right)+c_{\alpha _1}^2 m_{H_2}^2 s_{\theta }^2}{v^2 s_{\beta }^2}+\\ \notag
&\frac{c_{\theta }^2 m_h^2 s_{\alpha _2}^2-\frac{m _{12}^2 c_{\gamma }}{t_{\beta }}+m_{H_1}^2 s_{\alpha _1}^2 s_{\alpha _2}^2 s_{\theta }^2-\frac{m _{23}^2 s_{\gamma }}{t_{\beta }}}{v^2 s_{\beta }^2}\\
\lambda_3&=\frac{-c_{\alpha _2}^2 c_{\theta }^2 m_h^2 s_{\alpha _1}^2-2 c_{\alpha _2} c_{\theta } m_h^2 s_{\alpha _1} s_{\alpha _2} s_{\theta }+c_{\alpha _1}^2 \left(-c_{\theta }^2\right) m_{H_2}^2-c_{\alpha _2}^2 m_{H_1}^2 s_{\theta }^2}{v^2 c_{\beta }^2 s_{\gamma }^2}+\\ \notag
&-\frac{-c_{\theta }^2 m_{H_1}^2 s_{\alpha _1}^2 s_{\alpha _2}^2+2 c_{\alpha _2} c_{\theta } m_{H_1}^2 s_{\alpha _1} s_{\alpha _2} s_{\theta }-m_h^2 s_{\alpha _2}^2 s_{\theta }^2+\frac{m _{23}^2 t_{\beta }}{s_{\gamma }}+\frac{m _{13}^2}{t_{\gamma }}}{v^2 c_{\beta }^2 s_{\gamma }^2}\\  
\kappa_1&=\frac{1}{2 v^2 s_{\gamma }}\left(4 c_{\theta _2}^2 m_{H_1{}^\pm}^2 s_{\gamma }-\frac{2 c_{\alpha _1}^2 c_{\theta } m_{H_2}^2 s_{\theta }}{c_{\beta } s_{\beta }}+\frac{c_{\alpha _2}^2 s_{2 \theta } \left(c_{2\alpha _1} m_h^2-m_h^2+2 m_{H_1}^2\right)}{2 c_{\beta } s_{\beta }}-\right. \\ \notag
&\frac{c_{\alpha _2} s_{\alpha _1} s_{\alpha _2} \left(c_{2 \theta } \left(m_{H_1}^2-m_h^2\right)+m_h^2+m_{H_1}^2\right)}{c_{\beta } s_{\beta }}+\frac{m_h^2 s_{\alpha _2}^2 s_{2 \theta }}{c_{\beta } s_{\beta }}-\frac{2 c_{\theta } m_{H_1}^2 s_{\alpha _1}^2 s_{\alpha _2}^2 s_{\theta }}{c_{\beta } s_{\beta }}+\frac{c_{\theta }^2 m_h^2 s_{\alpha _1} s_{2 \alpha _2}}{c_{\beta } s_{\beta }}+ \\ \notag
&\frac{m_{H_1}^2 s_{\alpha _1} s_{2 \alpha _2} s_{\theta }^2}{c_{\beta } s_{\beta }}+\frac{m_{H_1{}^\pm}^2 s_{\theta _2}}{c_{\gamma } s_{\beta }}-\frac{m_{H_2{}^\pm}^2 s_{\theta _2}}{c_{\gamma } s_{\beta }}+4 m_{H_2{}^\pm}^2 s_{\gamma } s_{\theta _2}^2-\frac{m_{H_1{}^\pm}^2 s_{\gamma } s_{\theta _2} t_{\gamma }}{s_{\beta }}+ \\ \notag
&\left.\frac{c_{\gamma } \left(\left(m_{H_1{}^\pm}^2-m_{H_2{}^\pm}^2\right) s_{\theta _2}-\frac{2 m_{12}^2 s_{\gamma }}{c_{\beta }}\right)}{s_{\beta }}-\frac{2 m _{23}^2 c_{\gamma }^2}{c_{\beta } s_{\beta }}-\frac{2 m _{23}^2 s_{\gamma }^2}{c_{\beta } s_{\beta }}+\frac{2 m _{12}^2 t_{\gamma }}{c_{\beta } s_{\beta }}-\frac{2 m _{12}^2 s_{\gamma }^2 t_{\gamma }}{c_{\beta } s_{\beta }}+\frac{m_{H_2{}^\pm}^2 s_{\gamma } s_{\theta _2} t_{\gamma }}{s_{\beta }}\right)\\
&\rho_1=\frac{1}{2 v^2}\left(4 c_{\theta _2}^2 m_{H_1{}^\pm}^2-\frac{2 c_{\alpha _1} c_{\alpha _2}^2 m_h^2 s_{\alpha _1} s_{\theta }}{c_{\beta } c_{\gamma } s_{\beta }}+\frac{m_{H_2}^2 s_{2 \alpha _1} s_{\theta }}{c_{\beta } c_{\gamma } s_{\beta }}-\frac{2 c_{\alpha _1} c_{\alpha _2} c_{\theta } m_{H_1}^2 s_{\alpha _2}}{c_{\beta } c_{\gamma } s_{\beta }}-\frac{2 c_{\alpha _1} m_{H_1}^2 s_{\alpha _1} s_{\alpha _2}^2 s_{\theta }}{c_{\beta } c_{\gamma } s_{\beta }}+\right. \\ \notag
&\frac{c_{\alpha _1} c_{\theta } m_h^2 s_{2 \alpha _2}}{c_{\beta } c_{\gamma } s_{\beta }}-\frac{m_{H_1{}^\pm}^2 s_{\theta _2}}{c_{\gamma } s_{\beta } s_{\gamma }}+\frac{m_{H_2{}^\pm}^2 s_{\theta _2}}{c_{\gamma } s_{\beta } s_{\gamma }}+4 m_{H_2{}^\pm}^2 s_{\theta _2}^2+\frac{m_{H_1{}^\pm}^2 s_{\theta _2}}{s_{\beta } t_{\gamma }}-\frac{m_{H_2{}^\pm}^2 s_{\theta _2}}{s_{\beta } t_{\gamma }}-\\ \notag
&\left.\frac{2 m _{23}^2}{c_{\beta } s_{\beta } s_{\gamma }}-\frac{2 m _{23}^2 s_{\gamma }}{c_{\beta } s_{\beta }}-\frac{2 m _{12}^2 s_{\gamma } t_{\gamma }}{c_{\beta } s_{\beta }}-\frac{2 \left(m _{23}^2 c_{\gamma }+m _{12}^2 s_{\gamma }\right)}{c_{\beta } s_{\beta } t_{\gamma }}+\frac{m_{\left(H_1\right){}^+}^2 s_{\theta _2} t_{\gamma }}{s_{\beta }}+\frac{m_{\left(H_2\right){}^+}^2 s_{\theta _2} t_{\gamma }}{s_{\beta }}\right)\\ 
\sigma_1&=\frac{1}{8 v^2 c_{\beta }}\left(\frac{6 \left(c_{2 \beta }-2\right) c_{\theta _2}^2 \left(m_{H_1{}^\pm}^2-m_{H_2{}^\pm}^2\right)}{c_{\beta }}-\frac{8 c_{2 \gamma } \left(m_{H_1{}^\pm}^2-m_{H_2{}^\pm}^2\right) s_{\theta _2} t_{\beta }}{c_{\gamma } s_{\gamma }}-\right. \\ \notag
&\frac{1}{c_{\beta } c_{\gamma } s_{\gamma }}\left(-c_{\gamma } s_{\gamma } \left(c_{2 \beta } \left(c_{2 \theta } \left(m_{H_1{}^\pm}^2-m_{H_2{}^\pm}^2\right)+m_{\left(H_1\right){}^+}^2+7 m_{H_2{}^\pm}^2\right)+4 \left(m_{\left(H_1\right){}^+}^2+m_{H_2{}^\pm}^2\right)\right)+\right. \\ \notag
&8 c_{\alpha _1} \left(c_{\alpha _2} s_{\alpha _2} s_{\theta } \left(m_{H_1}^2-m_h^2\right)+c_{\theta } s_{\alpha _1} \left(-c_{\alpha _2}^2 m_h^2-m_{H_1}^2 s_{\alpha _2}^2+m_{H_2}^2\right)\right)-6 m_{H_1{}^\pm}^2 s_{2 \gamma } s_{\theta _2}^2+\\ \notag
&\left.\left.16 m _{13}^2 c_{\gamma }^2+6 m_{H_2{}^\pm}^2 s_{2 \gamma } s_{\theta _2}^2-8 m _{13}^2+4 m _{13}^2 s_{\gamma }^4+12 m _{13}^2 s_{\gamma }^2+m _{13}^2 s_{2 \gamma }^2\right)\right)
\end{align}

\section{Coefficients  of  $h H^{+}_{i} H^{-}_{i}$ Terms in  the Potential } \label{app-coeff}
The coefficients of triple scalar couplings in the potential are explicitly given as
\begin{align}
a=&\frac{-1}{2 v}\left(c_{\alpha _1} c_{\alpha _2} s_{\beta } \left(c_{\alpha _1} c_{\theta } s_{2 \alpha _2}-c_{\alpha _2}^2 s_{2 \alpha _1} s_{\theta }\right) \left(\frac{c_{\theta _2}}{t_{\beta }}+s_{\gamma } s_{\theta _2}\right){}^2-\right.\\ \notag
&s_{\beta } \left(c_{\alpha _2} c_{\theta } s_{\alpha _1}+s_{\alpha _2} s_{\theta }\right) \left(c_{\alpha _2}^2 s_{\alpha _1}^2 s_{2 \theta }-c_{2 \theta } s_{2 \alpha _2} s_{\alpha _1}-s_{\alpha _2}^2 s_{2 \theta }\right) \left(\frac{c_{\theta _2}}{t_{\beta }}+s_{\gamma } s_{\theta _2}\right){}^2-\\ \notag
&2 s_{\beta } \left(c_{\alpha _2} s_{\alpha _1} s_{\theta }-c_{\theta } s_{\alpha _2}\right) \left(c_{\alpha _2}^2 s_{\alpha _1}^2 s_{\theta }^2+c_{\theta } \left(c_{\theta } s_{\alpha _2}^2-s_{\alpha _1} s_{2 \alpha _2} s_{\theta }\right)\right) \left(\frac{c_{\theta _2}}{t_{\beta }}+s_{\gamma } s_{\theta _2}\right){}^2+\\ \notag
&2 c_{\alpha _1}^3 c_{\alpha _2}^3 c_{\gamma } s_{\beta } t_{\beta }+c_{\gamma } s_{\beta } t_{\beta } \left(c_{\alpha _2} s_{\alpha _1} s_{\theta }-c_{\theta } s_{\alpha _2}\right) \left(c_{\alpha _2}^2 s_{2 \alpha _1} s_{\theta }-c_{\alpha _1} c_{\theta } s_{2 \alpha _2}\right)+\\ \notag
&c_{\gamma } s_{\beta } t_{\beta } \left(c_{\alpha _2} c_{\theta } s_{\alpha _1}+s_{\alpha _2} s_{\theta }\right) \left(c_{\alpha _2}^2 c_{\theta } s_{2 \alpha _1}+c_{\alpha _1} s_{2 \alpha _2} s_{\theta }\right)+\frac{c_{\alpha _1} c_{\alpha _2} c_{\beta } \left(c_{\alpha _2}^2 c_{\theta } s_{2 \alpha _1}+c_{\alpha _1} s_{2 \alpha _2} s_{\theta }\right) \left(s_{\theta _2}-c_{\theta _2} s_{\gamma } t_{\beta }\right){}^2}{s_{\gamma }}+\\ \notag
&\frac{c_{\beta } \left(c_{\alpha _2} s_{\alpha _1} s_{\theta }-c_{\theta } s_{\alpha _2}\right) \left(c_{\alpha _2}^2 s_{\alpha _1}^2 s_{2 \theta }-c_{2 \theta } s_{2 \alpha _2} s_{\alpha _1}-s_{\alpha _2}^2 s_{2 \theta }\right) \left(s_{\theta _2}-c_{\theta _2} s_{\gamma } t_{\beta }\right){}^2}{s_{\gamma }}+\\ \notag
&\left.\frac{2 c_{\beta } \left(c_{\alpha _2} c_{\theta } s_{\alpha _1}+s_{\alpha _2} s_{\theta }\right) \left(c_{\alpha _2}^2 c_{\theta }^2 s_{\alpha _1}^2+c_{\theta } s_{2 \alpha _2} s_{\alpha _1} s_{\theta }+s_{\alpha _2}^2 s_{\theta }^2\right) \left(s_{\theta _2}-c_{\theta _2} s_{\gamma } t_{\beta }\right){}^2}{s_{\gamma }}\right)\\ 
b=&\frac{1}{4 v}\left(-\frac{4 \left(c_{\alpha _2} c_{\theta } s_{\alpha _1}+s_{\alpha _2} s_{\theta }\right) \left(c_{\beta } c_{\theta _2}+s_{\beta } s_{\gamma } s_{\theta _2}\right){}^2 \left(2 c_{\theta _2}^2 s_{\beta } s_{\gamma }+c_{\gamma } s_{2 \theta _2}\right)}{t_{\beta }}+\right.\\ \notag
&c_{\gamma }^2 s_{\beta }^2 \left(c_{\alpha _2} c_{\theta } s_{\alpha _1}+s_{\alpha _2} s_{\theta }\right) \left(\frac{4 c_{2 \gamma } s_{2 \theta _2} t_{\beta }}{c_{\gamma }}-\frac{2 s_{\gamma } \left(\left(2 c_{2 \beta }-3\right) c_{\theta _2}^2+3 s_{\theta _2}^2+1\right)}{c_{\beta }}\right)+\\ \notag
&c_{\gamma } s_{\beta } \left(c_{\gamma } \left(c_{\alpha _2} c_{\theta } s_{\alpha _1}+s_{\alpha _2} s_{\theta }\right)+c_{\alpha _1} c_{\alpha _2} s_{\gamma }\right) \left(c_{\beta } s_{\theta _2}-c_{\theta _2} s_{\beta } s_{\gamma }\right) \left(\frac{\left(6-4 c_{2 \beta }\right) c_{\theta _2}^2+3 c_{2 \theta _2}-5}{c_{\beta }}+\frac{4 c_{2 \gamma } s_{2 \theta _2} t_{\beta }}{c_{\gamma } s_{\gamma }}\right)+\\ \notag
&4 \left(c_{\alpha _2} s_{\alpha _1} s_{\theta }-c_{\theta } s_{\alpha _2}\right) \left(c_{\theta _2} s_{\beta } s_{\gamma }-c_{\beta } s_{\theta _2}\right){}^2 \left(2 c_{\theta _2}^2 s_{\beta }+\frac{s_{2 \theta _2}}{t_{\gamma }}\right)+4\left(c_{\beta } s_{\gamma } \left(c_{\theta } s_{\alpha _2}-c_{\alpha _2} s_{\alpha _1} s_{\theta }\right)+\right.\\ \notag
& \left.+s_{\beta } \left(c_{\alpha _2} c_{\theta } s_{\alpha _1}+s_{\alpha _2} s_{\theta }\right)\right)(c_{\beta } s_{\theta _2}-c_{\theta _2} s_{\beta } s_{\gamma })\left(c_{\beta } c_{\theta _2}+s_{\beta } s_{\gamma } s_{\theta _2}\right) \left(2 c_{\theta _2}^2+\frac{s_{2 \theta _2}}{s_{\beta } t_{\gamma }}\right)+\\ \notag
&\frac{c_{\alpha _1} c_{\alpha _2} \left(c_{\theta _2} s_{\beta } s_{\gamma }-c_{\beta } s_{\theta _2}\right){}^2 \left(\frac{4 c_{2 \gamma } s_{2 \theta _2} t_{\beta }}{c_{\gamma }}-\frac{2 s_{\gamma } \left(\left(2 c_{2 \beta }-3\right) c_{\theta _2}^2+3 s_{\theta _2}^2+1\right)}{c_{\beta }}\right)}{t_{\gamma }}+\\ \notag
&4 c_{\gamma }^2 s_{\beta }^2 \left(c_{\alpha _2} s_{\alpha _1} s_{\theta }-c_{\theta } s_{\alpha _2}\right) \left(2 c_{\theta _2}^2 s_{\beta }-s_{2 \theta _2} t_{\gamma }\right)-\frac{4 c_{\alpha _1} c_{\alpha _2} c_{\gamma } \left(c_{\beta } c_{\theta _2}+s_{\beta } s_{\gamma } s_{\theta _2}\right){}^2 \left(2 c_{\theta _2}^2 s_{\beta }-s_{2 \theta _2} t_{\gamma }\right)}{t_{\beta }}+\\ \notag
&\left.c_{\gamma } s_{\beta } \left(c_{\beta } c_{\theta _2}+s_{\beta } s_{\gamma } s_{\theta _2}\right) \left(c_{\beta } c_{\gamma } \left(c_{\theta } s_{\alpha _2}-c_{\alpha _2} s_{\alpha _1} s_{\theta }\right)+c_{\alpha _1} c_{\alpha _2} s_{\beta }\right) \left(-8 c_{\theta _2}^2-\frac{4 c_{\theta _2} s_{\theta _2}}{s_{\beta } t_{\gamma }}+\frac{2 s_{2 \theta _2} \left(\frac{1}{c_{\gamma } s_{\gamma }}+t_{\gamma }\right)}{s_{\beta }}\right)\right)
\end{align}
\begin{align}
c=&\frac{1}{8 v}\left(-\frac{8 \left(c_{\alpha _2} c_{\theta } s_{\alpha _1}+s_{\alpha _2} s_{\theta }\right) \left(c_{\beta } c_{\theta _2}+s_{\beta } s_{\gamma } s_{\theta _2}\right){}^2 \left(2 s_{\beta } s_{\gamma } s_{\theta _2}^2-c_{\gamma } s_{2 \theta _2}\right)}{t_{\beta }}+\right.\\ \notag
&2 c_{\gamma }^2 s_{\beta }^2 \left(c_{\alpha _2} c_{\theta } s_{\alpha _1}+s_{\alpha _2} s_{\theta }\right) \left(-\frac{s_{\gamma } \left(6 c_{\theta _2}^2+\left(4 c_{2 \beta }-6\right) s_{\theta _2}^2+2\right)}{c_{\beta }}-\frac{4 c_{2 \gamma } s_{2 \theta _2} t_{\beta }}{c_{\gamma }}\right)+\\ \notag
&c_{\gamma } s_{\beta } \left(c_{\gamma } \left(c_{\alpha _2} c_{\theta } s_{\alpha _1}+s_{\alpha _2} s_{\theta }\right)+c_{\alpha _1} c_{\alpha _2} s_{\gamma }\right)( c_{\beta } s_{\theta _2}-c_{\theta _2} s_{\beta } s_{\gamma })\left(\frac{2 \left(c_{\beta -\theta _2}+c_{\beta +\theta _2}-2 c_{2 \beta }-6 c_{2 \theta _2}-2\right)}{c_{\beta }}-\right.\\ \notag
&\left.\frac{8 c_{2 \gamma } s_{2 \theta _2} t_{\beta }}{c_{\gamma } s_{\gamma }}\right)+8 \left(c_{\alpha _2} s_{\alpha _1} s_{\theta }-c_{\theta } s_{\alpha _2}\right) \left(c_{\theta _2} s_{\beta } s_{\gamma }-c_{\beta } s_{\theta _2}\right){}^2 \left(2 s_{\beta } s_{\theta _2}^2-\frac{s_{2 \theta _2}}{t_{\gamma }}\right)+\\ \notag
&4 s_{\beta }^2 \left(c_{\alpha _2} s_{\alpha _1} s_{\theta }-c_{\theta } s_{\alpha _2}\right) \left(c_{\gamma } s_{\gamma } s_{2 \theta _2}+c_{\gamma }^2 \left(4 s_{\beta } s_{\theta _2}^2-\frac{s_{2 \theta _2}}{t_{\gamma }}\right)+\frac{s_{2 \theta _2}}{t_{\gamma }}\right)+\\ \notag
&\frac{2 c_{\alpha _1} c_{\alpha _2} \left(c_{\theta _2} s_{\beta } s_{\gamma }-c_{\beta } s_{\theta _2}\right){}^2 \left(-\frac{s_{\gamma } \left(6 c_{\theta _2}^2+\left(4 c_{2 \beta }-6\right) s_{\theta _2}^2+2\right)}{c_{\beta }}-\frac{4 c_{2 \gamma } s_{2 \theta _2} t_{\beta }}{c_{\gamma }}\right)}{t_{\gamma }}-\\ \notag
&2\left(c_{\beta } s_{\gamma } \left(c_{\theta } s_{\alpha _2}-c_{\alpha _2} s_{\alpha _1} s_{\theta }\right)+\right.\left.s_{\beta } \left(c_{\alpha _2} c_{\theta } s_{\alpha _1}+s_{\alpha _2} s_{\theta }\right)\right)(c_{\beta } s_{\theta _2}-c_{\theta _2} s_{\beta } s_{\gamma }) (c_{\beta } c_{\theta _2}+s_{\beta } s_{\gamma } s_{\theta _2}) \\ \notag
&\left(-8 s_{\theta _2}^2+\frac{2 s_{2 \theta _2} \left(\frac{1}{c_{\gamma } s_{\gamma }}+\frac{1}{t_{\gamma }}\right)}{s_{\beta }}-\frac{4 c_{\theta _2} s_{\theta _2} t_{\gamma }}{s_{\beta }}\right)+2 c_{\gamma } s_{\beta } \left(c_{\beta } c_{\gamma } \left(c_{\theta } s_{\alpha _2}-c_{\alpha _2} s_{\alpha _1} s_{\theta }\right)+c_{\alpha _1} c_{\alpha _2} s_{\beta }\right)\\ \notag
&\left(c_{\beta } c_{\theta _2}+s_{\beta } s_{\gamma } s_{\theta _2}\right) \left(-\frac{4 c_{\theta _2} \left(\frac{1}{s_{\gamma }^2}+1\right) s_{\theta _2} t_{\gamma }}{s_{\beta }}-8 s_{\theta _2}^2+\frac{2 s_{2 \theta _2}}{s_{\beta } t_{\gamma }}\right)-\\ \notag
&\left.\frac{8 c_{\alpha _1} c_{\alpha _2} c_{\gamma } \left(c_{\beta } c_{\theta _2}+s_{\beta } s_{\gamma } s_{\theta _2}\right){}^2 \left(2 s_{\beta } s_{\theta _2}^2+s_{2 \theta _2} t_{\gamma }\right)}{t_{\beta }}\right)\\ 
d=&-\frac{-c_{\alpha _1} c_{\alpha _2} s_{\beta } \left(\frac{c_{\theta _2}}{t_{\beta }}+s_{\gamma } s_{\theta _2}\right){}^2+c_{\beta } c_{\gamma } \left(c_{\alpha _2} s_{\alpha _1} s_{\theta }-c_{\theta } s_{\alpha _2}\right) \left(\frac{c_{\theta _2}}{t_{\beta }}+s_{\gamma } s_{\theta _2}\right){}^2+c_{\gamma } s_{\beta } t_{\beta } \left(c_{\alpha _2} s_{\alpha _1} s_{\theta }-c_{\theta } s_{\alpha _2}\right)}{v}\\ \notag
&+\frac{c_{\alpha _1} c_{\alpha _2} s_{\beta } t_{\beta }^2+2 c_{\beta } \left(c_{\theta _2}+s_{\gamma } s_{\theta _2} t_{\beta }\right) \left(c_{\gamma } \left(c_{\theta } s_{\alpha _2}-c_{\alpha _2} s_{\alpha _1} s_{\theta }\right)+c_{\alpha _1} c_{\alpha _2} t_{\beta }\right)}{v}\\ 
e=&\frac{c_{\beta } \left(s_{\theta _2}+2 s_{\gamma } s_{\frac{\theta _2}{2}}^2 t_{\beta }\right){}^2 \left(c_{\alpha _1} c_{\alpha _2}+\frac{c_{\alpha _2} c_{\theta } s_{\alpha _1}+s_{\alpha _2} s_{\theta }}{t_{\gamma }}\right)}{v s_{\gamma }} \\
f=&\frac{\left(\frac{c_{\theta _2}}{c_{\beta }}-\frac{c_{\gamma } s_{\beta } s_{\theta _2}}{t_{\gamma }}\right){}^2 \left(c_{\theta } \left(c_{\alpha _2} s_{\alpha _1}+\frac{s_{\alpha _2} s_{\gamma }}{t_{\beta }}\right)+s_{\theta } \left(s_{\alpha _2}-\frac{c_{\alpha _2} s_{\alpha _1} s_{\gamma }}{t_{\beta }}\right)\right)}{v s_{\beta }}\\
a_1=&-\frac{\frac{c_{\gamma }^2 s_{\theta _2}^2 \left(c_{\theta } s_{\alpha _2}-c_{\alpha _2} s_{\alpha _1} s_{\theta }\right)}{s_{\beta }}+\frac{c_{\gamma } c_{\theta _2}^2 \left(c_{\alpha _2} c_{\theta } s_{\alpha _1}+s_{\alpha _2} s_{\theta }\right)}{c_{\beta } t_{\gamma }}+\frac{c_{\alpha _1} c_{\alpha _2} s_{\gamma } t_{\gamma }}{c_{\beta }}}{v}
\end{align}
\newpage
\begin{align}
b_1&=-\frac{1}{4 v}\left(4 c_{\beta } c_{\gamma }^2 s_{\theta _2}^2 \left(c_{\alpha _2} c_{\theta } s_{\alpha _1}+s_{\alpha _2} s_{\theta }\right) \left(\frac{c_{\gamma } s_{2 \theta _2}}{s_{\beta }}+2 c_{\theta _2}^2 s_{\gamma }\right)-\right. \\  \notag
&s_{\gamma }^2 \left(c_{\alpha _2} c_{\theta } s_{\alpha _1}+s_{\alpha _2} s_{\theta }\right) \left(\frac{4 c_{2 \gamma } s_{2 \theta _2} t_{\beta }}{c_{\gamma }}-\frac{2 s_{\gamma } \left(\left(2 c_{2 \beta }-3\right) c_{\theta _2}^2+3 s_{\theta _2}^2+1\right)}{c_{\beta }}\right)-\\ \notag
&c_{\gamma } c_{\theta _2} s_{\gamma } \left(c_{\gamma } \left(c_{\alpha _2} c_{\theta } s_{\alpha _1}+s_{\alpha _2} s_{\theta }\right)+c_{\alpha _1} c_{\alpha _2} s_{\gamma }\right) \left(\frac{\left(6-4 c_{2 \beta }\right) c_{\theta _2}^2+3 c_{2 \theta _2}-5}{c_{\beta }}+\frac{4 c_{2 \gamma } s_{2 \theta _2} t_{\beta }}{c_{\gamma } s_{\gamma }}\right)-\\ \notag
&2 c_{\theta _2}^2 \left(c_{\alpha _2} s_{\alpha _1} s_{\theta }-c_{\theta } s_{\alpha _2}\right) \left(-2 c_{\gamma } c_{\theta _2} s_{\gamma } s_{\theta _2}+c_{\gamma }^2 \left(4 c_{\theta _2}^2 s_{\beta }+\frac{s_{2 \theta _2}}{t_{\gamma }}\right)+\frac{s_{2 \theta _2}}{t_{\gamma }}\right)+\\ \notag
&\frac{4 c_{\gamma } c_{\theta _2} s_{\theta _2} \left(2 c_{\theta _2}^2 s_{\beta } s_{\gamma }+c_{\gamma } s_{2 \theta _2}\right) \left(s_{\theta } \left(s_{\alpha _2} s_{\beta }-c_{\alpha _2} c_{\beta } s_{\alpha _1} s_{\gamma }\right)+c_{\theta } \left(c_{\beta } s_{\alpha _2} s_{\gamma }+c_{\alpha _2} s_{\alpha _1} s_{\beta }\right)\right)}{s_{\beta } t_{\gamma }}-\\ \notag
&\frac{c_{\alpha _1} c_{\alpha _2} c_{\gamma }^2 c_{\theta _2}^2 \left(\frac{4 c_{2 \gamma } s_{2 \theta _2} t_{\beta }}{c_{\gamma }}-\frac{2 s_{\gamma } \left(\left(2 c_{2 \beta }-3\right) c_{\theta _2}^2+3 s_{\theta _2}^2+1\right)}{c_{\beta }}\right)}{t_{\gamma }}+2 \left(c_{\alpha _2} s_{\alpha _1} s_{\theta }-c_{\theta } s_{\alpha _2}\right)\left(-4 c_{\gamma } c_{\theta _2}^2 s_{\beta } s_{\gamma }-\right.\\  \notag
&c_{\gamma }^2 s_{2 \theta _2}+ \left.\left(s_{\gamma }^2+1\right) s_{2 \theta _2}\right)t_{\gamma }+4 c_{\alpha _1} c_{\alpha _2} c_{\beta } c_{\gamma }^3 s_{\theta _2}^2 \left(2 c_{\theta _2}^2-\frac{s_{2 \theta _2} t_{\gamma }}{s_{\beta }}\right)+c_{\gamma }s_{\gamma }\left(c_{\alpha _1} c_{\alpha _2} s_{\beta }+\right.\\ \notag
&\left.\left.c_{\beta } c_{\gamma } \left(c_{\theta } s_{\alpha _2}-c_{\alpha _2} s_{\alpha _1} s_{\theta }\right)\right)s_{\theta _2}\left( -8 c_{\theta _2}^2-\frac{4 c_{\theta _2} s_{\theta _2}}{s_{\beta } t_{\gamma }}+\frac{2 s_{2 \theta _2} \left(\frac{1}{c_{\gamma } s_{\gamma }}+t_{\gamma }\right)}{s_{\beta }}\right)\right) \\ 
c_1=&\frac{1}{8 v}\left(8 c_{\beta } c_{\gamma }^2 s_{\theta _2}^2 \left(c_{\alpha _2} c_{\theta } s_{\alpha _1}+s_{\alpha _2} s_{\theta }\right) \left(\frac{c_{\gamma } s_{2 \theta _2}}{s_{\beta }}-2 s_{\gamma } s_{\theta _2}^2\right)+\right.\\ \notag
&2 s_{\gamma }^2 \left(c_{\alpha _2} c_{\theta } s_{\alpha _1}+s_{\alpha _2} s_{\theta }\right) \left(-\frac{s_{\gamma } \left(6 c_{\theta _2}^2+\left(4 c_{2 \beta }-6\right) s_{\theta _2}^2+2\right)}{c_{\beta }}-\frac{4 c_{2 \gamma } s_{2 \theta _2} t_{\beta }}{c_{\gamma }}\right)+\\ \notag
&c_{\gamma } c_{\theta _2} s_{\gamma } \left(c_{\gamma } \left(c_{\alpha _2} c_{\theta } s_{\alpha _1}+s_{\alpha _2} s_{\theta }\right)+c_{\alpha _1} c_{\alpha _2} s_{\gamma }\right) \left(\frac{2 \left(c_{\beta -\theta _2}+c_{\beta +\theta _2}-2 c_{2 \beta }-6 c_{2 \theta _2}-2\right)}{c_{\beta }}-\frac{8 c_{2 \gamma } s_{2 \theta _2} t_{\beta }}{c_{\gamma } s_{\gamma }}\right)-\\ \notag
&4 c_{\theta _2}^2 \left(c_{\alpha _2} s_{\alpha _1} s_{\theta }-c_{\theta } s_{\alpha _2}\right) \left(-2 c_{\gamma } c_{\theta _2} s_{\gamma } s_{\theta _2}+c_{\gamma }^2 \left(\frac{s_{2 \theta _2}}{t_{\gamma }}-4 s_{\beta } s_{\theta _2}^2\right)+\frac{s_{2 \theta _2}}{t_{\gamma }}\right)+\\ \notag
&\frac{2 c_{\alpha _1} c_{\alpha _2} c_{\gamma }^2 c_{\theta _2}^2 \left(-\frac{s_{\gamma } \left(6 c_{\theta _2}^2+\left(4 c_{2 \beta }-6\right) s_{\theta _2}^2+2\right)}{c_{\beta }}-\frac{4 c_{2 \gamma } s_{2 \theta _2} t_{\beta }}{c_{\gamma }}\right)}{t_{\gamma }}+4 \left(c_{\alpha _2} s_{\alpha _1} s_{\theta }-c_{\theta } s_{\alpha _2}\right)\left(4 c_{\gamma } s_{\beta } s_{\gamma } s_{\theta _2}^2-c_{\gamma }^2 s_{2 \theta _2}+\right.\\ \notag
&\left.\left(s_{\gamma }^2+1\right) s_{2 \theta _2}\right)t_{\gamma }+2c_{\gamma }^2c_{\theta _2}\left(c_{\beta } s_{\gamma } \left(c_{\theta } s_{\alpha _2}-c_{\alpha _2} s_{\alpha _1} s_{\theta }\right)+\right. \left.s_{\beta } \left(c_{\alpha _2} c_{\theta } s_{\alpha _1}+s_{\alpha _2} s_{\theta }\right)\right)s_{\theta _2}\\ \notag
&\left( -\frac{4 c_{\theta _2} s_{\theta _2} t_{\gamma }}{s_{\beta }}+\frac{2 s_{2 \theta _2} \left(\frac{1}{c_{\gamma } s_{\gamma }}+\frac{1}{t_{\gamma }}\right)}{s_{\beta }}-8 s_{\theta _2}^2\right) \\ \notag
&-2 c_{\gamma } s_{\gamma } s_{\theta _2} \left(c_{\beta } c_{\gamma } \left(c_{\theta } s_{\alpha _2}-c_{\alpha _2} s_{\alpha _1} s_{\theta }\right)+c_{\alpha _1} c_{\alpha _2} s_{\beta }\right) \left(-\frac{4 c_{\theta _2} \left(\frac{1}{s_{\gamma }^2}+1\right) s_{\theta _2} t_{\gamma }}{s_{\beta }}-8 s_{\theta _2}^2+\frac{2 s_{2 \theta _2}}{s_{\beta } t_{\gamma }}\right) \\ \notag
&\left.-8 c_{\alpha _1} c_{\alpha _2} c_{\beta } c_{\gamma }^3 s_{\theta _2}^2 \left(2 s_{\theta _2}^2+\frac{s_{2 \theta _2} t_{\gamma }}{s_{\beta }}\right)\right)\\ \notag
\end{align}
\newpage
\begin{align}
d_1=&\frac{\left(c_{\beta } c_{\gamma } \left(c_{\theta } s_{\alpha _2}-c_{\alpha _2} s_{\alpha _1} s_{\theta }\right)+c_{\alpha _1} c_{\alpha _2} s_{\beta }\right) \left(\frac{8 c_{\gamma } s_{\theta _2}}{s_{\beta }}-\frac{8 t_{\gamma }}{c_{\beta }}\right){}^2}{64 v}\\ 
e_1=&\frac{s_{\gamma } \left(\frac{1}{c_{\gamma } s_{\gamma }}+\frac{2 c_{\theta _2}-1}{t_{\gamma }}+t_{\gamma }\right){}^2 \left(c_{\alpha _1} c_{\alpha _2}+\frac{c_{\alpha _2} c_{\theta } s_{\alpha _1}+s_{\alpha _2} s_{\theta }}{t_{\gamma }}\right)}{4 v c_{\beta }}\\
f_1=&\frac{\left(s_{\theta } \left(s_{\alpha _2} s_{\beta }-c_{\alpha _2} c_{\beta } s_{\alpha _1} s_{\gamma }\right)+c_{\theta } \left(c_{\beta } s_{\alpha _2} s_{\gamma }+c_{\alpha _2} s_{\alpha _1} s_{\beta }\right)\right) \left(\frac{c_{\gamma } s_{\theta _2}}{s_{\beta }}+\frac{c_{\theta _2}}{c_{\beta } t_{\gamma }}\right){}^2}{v}
\end{align}

\newpage

\vspace*{-4mm}
\section{Definitions of the  One-Loop Factors}\label{loop function} 
The detailed analytic formulas of the one-loop factors  are given as follows \cite{djouadi, djouadi1}:
\begin{align}
A_{1/2}(\tau)&=2[\tau+(\tau-1)f(\tau)]\tau^{-2} \\ \notag
A_{1}(\tau)&= - [2\tau^{2}+3\tau +3(2\tau-1)f(\tau)] \tau^{-2} \\ \notag
A_{0}(\tau)&=-[\tau-f(\tau)]\tau^{-2}
\end{align}
The function $f(\tau)$ is given by
\begin{align}
f(\tau)=\begin{cases}
    \text{arcsin}^2 \sqrt{\tau} &  \tau\leq 1\\
    -\frac{1}{4}\left[ \text{log}\frac{1+\sqrt{1-\tau^{-1}}}{1-\sqrt{1-\tau^{-1}}} -i\pi\right] & \tau>1.
  \end{cases} 
\end{align}

We used 
\begin{align*}
\tau_i=\frac{m_{\phi}^2}{4m_i^2}~ ~~\text{with} ~~i=f,W^{\pm},H^{\pm}_{1,2}. 
\end{align*}

Now $\tau_i=\frac{4m_{i}^2}{m_\phi^2}$, $ \lambda_i=\frac{4m_i^2}{m_Z^2}  $ and other loop factors

\begin{align}
A_{1/2}(\tau,\lambda)=&[I_1(\tau,\lambda)-I_2(\tau,\lambda)] \\ 
A_1(\tau, \lambda)=&c_W\left\lbrace 4\left( 3-\frac{s_W^2}{c_W^2}\right) I_2(\tau, \lambda)+\left[ \left( 1+\frac{2}{\tau}\right) \frac{s_W^2}{c_W^2}-\left( 5+\frac{2}{\tau}\right) \right]I_1(\tau, \lambda) \right\rbrace 
\end{align}
The functions $ I_1 $ and $ I_2 $ are
\begin{align}
I_1(\tau, \lambda)=&\frac{\tau \lambda}{2(\tau-\lambda)}+\frac{\tau^2 \lambda^2}{2(\tau-\lambda)^2}[f(\tau^{-1})-f(\lambda^{-1})]+\frac{\tau^2 \lambda}{(\tau-\lambda)^2}[g(\tau^{-1})-g(\lambda^{-1})] \\ 
I_2(\tau, \lambda)=&-\frac{\tau\lambda}{2(\tau-\lambda)}[f(\tau^{-1})-f(\lambda^{-1})]
\end{align}
The function $g(\tau)$ is written as
\begin{align}
g(\tau)=\begin{cases}
 \sqrt{\tau^{-1}-1}   ~\text{arcsin} \sqrt{\tau} &  \tau \geq1\\
    \frac{\sqrt{1-\tau^{-1}}}{2}\left[ \text{log}\frac{1+\sqrt{1-\tau^{-1}}}{1-\sqrt{1-\tau^{-1}}} -i\pi\right] & \tau<1.
  \end{cases}
\end{align}
\end{appendix}

\end{document}